\documentclass[twocolumn,a4paper,aps,amsfont,preprintnumbers,balancelastpage,showkeys,10pt]{revtex4-2}
\usepackage{amsmath}
\usepackage{amssymb}
\usepackage{amsfonts}
\usepackage{fontenc}
\usepackage[mathcal]{eucal}
\usepackage{bm}
\usepackage{lmodern}
\usepackage{epstopdf}
\usepackage{epsfig}
\usepackage{multirow}
\usepackage{dcolumn}
\usepackage{array}
\usepackage{booktabs}
\usepackage{indentfirst}
\newcommand{\tabitem}{~~\llap{\textbullet}~~}
\usepackage{tabularray}
\usepackage{rotating}
\usepackage[breaklinks]{hyperref}
\usepackage{makecell}
\usepackage{graphicx}
\usepackage{subfigure}
\DeclareMathAlphabet{\mathpzc}{OT1}{pzc}{m}{it}
\begin{document}
	\title{Neutron Star Core-Crust Transition and Crustal Moment of Inertia: \\ Systematic Implications of Higher-Order Symmetry Energy in the \\ Nuclear Equation of State}
	\author{W. M. Seif$^{1,2,3,(*)}$ and A. S. Hashem$^{1}$}
	\affiliation{\footnotesize $^{1}$Cairo University, Faculty of Science, Department of Physics, 12613 Giza, Egypt\\
	$^{2}$Joint Institute for Nuclear Research, 141980 Dubna, Russia\\$^{3}$The Academy of Scientific Research and Technology, 4262104 Cairo, Egypt\\
	$^{(*)}$\href{mailto:wseif@sci.cu.edu.eg}{{wseif@sci.cu.edu.eg}}}
	
\begin{abstract}
	We investigate how higher-order symmetry-energy coefficients of the equation of state (EOS) describing $npe\mu$ core matter impact key neutron star (NS) properties at its crust inner edge, its moment of inertia and corresponding crustal fraction, threshold conditions for direct Urca process, adiabatic index, and related structure observables. Our analysis employs a comprehensive set of CDM3Y-IVF1 equations of state, derived from the M3Y-Paris nucleon-nucleon interaction within a non-relativistic Hartree-Fock framework, covering a wide range of nuclear matter stiffness, from soft (${K}_{0}=150$ MeV) to extremely stiff (330 MeV) regimes. Our analysis reveals that the stiffer EOS characterized by higher isoscalar incompressibility (${K}_{0}$) and fourth-order symmetry-energy skewness (${Q}_{4}$) coefficients, coupled with diminished negative values of isoscalar ${Q}_{0}$ and ${H}_{0}$, and isovector ${K}_{4}$, ${K}_{\tau4}$, ${I}_{2,4}$, and ${G}_{2}$ coefficients, systematically increase the estimated core-crust transition pressure, density, and charged-particle ($p,e,\mu$) fractions, and their threshold values for direct Urca process, as well as the total NS moment of inertia and its crustal fraction, and the central adiabatic index. Simultaneously, they yield less central fraction of $p$, $e$, and $\mu$, and reduced fractional crust thickness and moment of inertia for the maximum NS mass supported by a certain EOS. Conversely, enhanced values of the isoscalar kurtosis ${I}_{0}$ and ${G}_{0}$ symmetry coefficients, and of the higher-order ${Q}_{2}$, ${H}_{2}$, ${K}_{6}$ and ${K}_{\tau6}$ coefficients, combined with diminished negative contributions from ${K}_{2}$ and ${K}_{\tau2}$, produce systematically opposite effects on the mentioned NS quantities, reversing their trends with ${K}_{0}$.
\end{abstract}	
	
	
\maketitle

\section{\label{sec:level1}Introduction}
\textit{Born from the violent collapse of massive stars in supernovae, neutron stars exhibit extremely high densities and extraordinary thermodynamic behaviors, making them natural cosmic laboratories for investigating nuclear matter (NM) under extreme conditions unattainable on Earth} \cite{LattimerPrakash2007}. The moment of inertia (MI) is one of the fundamental neutron star (NS) properties that quantifies the resistance to changes in rotation. It is a crucial component in understanding the structure and dynamics of slowly rotating NS. This quantity is influenced by the mass distribution within the NS, which is determined by the uncertain equation of state (EOS) of dense NM \cite{Ozeletal2016}. Pulsar glitches act as indirect indicator of the moment of inertia, offering insights into angular momentum exchanges between the superfluid core and crust \cite{AlparAnderson1984}. Meanwhile, gravitational wave (GW) signal detections by Advanced LIGO, Virgo, and KAGRA, coupled with X-ray observations from NICER, provide complementary constraints on stellar masses, radii, tidal deformability, crustal characteristics, MI and its crustal fraction, in addition to the related EOS \cite{Abbottetal2017,Abbottetal2020,AgulloCardoso2021,Milleretal2021,GittinsAndersson2020}, linking theoretical models to observational constraints. Additionally, radio timing studies refine our understanding of rotational dynamics and glitch mechanisms \cite{AntonopoulouHaskell2022,EspinozaAntonopoulou2021}. The MI is involved in many astrophysical processes, such as angular momentum transfer during the inspiral phase of binary systems \cite{FlanaganHinderer2008} and the stability of NS against rotational instabilities like the Chandrasekhar-Friedman-Schutz instability, which may drive continuous GWs emission \cite{AnderssonKokkotas1998}. Cutting-edge EOS frameworks, accounting for exotic phases such as hyperons or quark matter, have sharpened theoretical predictions of how the MI varies with mass and radius \cite{Weber2017,SteinerHempel2013}. Such combined theoretical and observational efforts deepen our understanding of NS structure and beyond. \par 
The solid crust of NSs is essential for understanding their rotational dynamics, thermal evolution, and observable emission mechanisms. The fraction of the MI belonging to the crust relative to the total moment of inertia, $\Delta I/I={I}_{\text{crust}}/{I}_{\text{total}}$, serves as a key probe of internal structure and constraining the matter and mass distributions within the NS. It provides valuable insights into the EOS of dense NM at extreme conditions and the progressive transitions from the superfluid core through the inner crust, comprising nuclear clusters permeated by neutron superfluidity, to the nuclear lattice structure of the outer crust. The crustal fraction of the MI is also essential for comprehending other astrophysical phenomena such as pulsar glitches, which result from sudden increases in spin frequency due to angular momentum transfer from the superfluid interior to the solid crust \cite{AlparAnderson1984}. Entrainment effects elevate the required crustal moment fraction to about six times initial estimates of 1.4\%, necessitating either anomalously thick crusts or core superfluid participation in glitch dynamics \cite{DutraLenzi2021}. Furthermore, the crustal fraction influences the resistance of NS against deformation instabilities, which could sustain persistent detectable GW emission \cite{UshomirskyCutler2000}. Observationally, X-ray bursts from accreting NSs provide constraints on the thickness and composition of their crusts \cite{Brownetal2002}, while the thermal relaxation and cooling behavior after transient heating events constraints both thermal properties and core-crust energy exchange \cite{PageGeppert2006}. The crustal fraction exhibits strong EOS dependence, where stiffer EOSs predict thicker crusts and larger $\Delta I/I$ \cite{SteinerHempel2013}. Exotic phases such as hyperons and quark matter may modify these quantities and also the conditions under which the direct Urca DU process becomes allowed. Modern EOS frameworks now incorporate considerations of phase transition dynamics and nuclear pasta phases \cite{Weber2017,Parmaretal2022,OyamatsuIida2020}. \par
The thermal evolution of NSs offers powerful diagnostics of their internal composition and structure, with the DU process standing out as one of the most efficient cooling pathways. This reaction channel entails the transformation of neutrons into protons and electrons, accompanied by the emission of electron neutrinos, through weak interaction of $\beta$ decay, permitted when the neutron Fermi level exceeds the combined Fermi level energies of protons and electrons \cite{LattimerYahil1989}. Unlike the modified Urca mechanism, the DU process occurs without intermediate baryons involved, yielding enhanced neutrino luminosity and accelerated thermal relaxation. The critical proton fraction needed for this process derived from the dense matter EOS, with stiffer NM favors the DU process and yields higher proton fractions at high densities \cite{SeifHashem2022}. X-ray spectral monitoring across isolated NSs and those in binary systems enable evolution timelines of surface temperature, providing insights into cooling history \cite{PageGeppert2006}. The rapid cooling observed in young NSs like Cassiopeia A has been attributed to DU process \cite{Shterninetal2011}. Parallel evidence comes from enhanced neutrino emission in low-mass X-ray binaries \cite{Brownetal2002}. \par
\textit{Composed of lattices of heavy nuclei in an electron sea, the crust bridges microscopic nuclear physics and macroscopic behavior. The crust-core boundary serves as a dynamic interface where ordered nuclear structures give way to quantum fluid behaviors}. This transition takes place at densities higher than the "drip density", where neutrons escape from nuclei, creating a neutron-rich environment of nucleons and leptons in $\beta$ equilibrium \cite{BaymBethe1971}. This transition is marked by substantial changes in composition, density, and pressure, which are highly sensitive to the EOS and the symmetry energy and its isovector density dependence. Modern theoretical frameworks like Br\"{u}ckner-Hartree-Fock (BHF) and relativistic mean-field approaches predict diverse EOS behaviors near this interface, with implications for GW signatures and inferred mass-radius curves. These models predict varying levels of stiffness in the EOS, which influence the structure and dynamics of NSs \cite{AkmalPandharipande1998}. Advanced computational techniques continue to develop our understanding of such complex systems \cite{Hebeleretal2013}, connecting ab initio predictions with empirical measurements. \par
Towards a realistic nuclear EOS, exact expressions for the EOS of asymmetric nuclear matter (ANM), and its energy and pressure, and also explicit analytical forms of its related symmetry coefficients, were derived using the Hartree-Fock (HF) approach employing the CDM3Y effective form of the semi-microscopic M3Y nucleon-nucleon (NN) potential, covering a broad stiffness range \cite{SeifHashem2023,Seif2011,KhoaOertzen1996}. These coefficients, analytically derived from higher-order derivatives of the binding and symmetry energy, provide full density and isospin-asymmetry description of the EOS, including higher-order isoscalar and isovector contributions. The validity of these coefficients was tested against constraints from nuclear data on finite nuclei and reactions, as well as astrophysical observations of NS \cite{SeifHashem2022,Seif2012,SeifHashem2021}. While quadratic terms adequately describe ANM at moderate densities less than twice the saturation density \cite{Seif2012}, it was shown that the higher-order symmetry-energy contributions up to eighth order significantly influence the proton fraction in $\beta$-stable NS matter, as well as the core-crust transition properties \cite{SeifBasu2014}. In prior work \cite{SeifHashem2025}, we explored how the density-isospin curvature, skewness, kurtosis, and higher-order isoscalar and isovector coefficients in the ANM EOS affect key NS structural properties, namely mass, radius, tidal deformability, Love number, compactness, core speed of sound, and surface redshift. In this work, the symmetry-energy coefficients have been constrained using observed NS masses, causality conditions on the speed of sound, the threshold proton fraction for DU processes, and NM pressure constraints from flow data in heavy-ion collisions. \par
In this study, we expand our analysis to explore the impact of higher-order symmetry energy coefficients on different key properties of slowly rotating NSs. Specifically, we examine their influence on the total MI and its crustal contribution, fractional crust thickness, DU cooling processes, transition properties at the core-crust interface, and the related adiabatic index, according to the stiffness of the employed CDM3Y EOSs. In the next section, we present the theoretical framework of the employed EOSs and their associated symmetry-energy coefficients, along with their computational formalism for deriving the NS properties under investigation. The numerical results and corresponding analysis are presented and discussed in Section \ref{sec:level3}. We conclude with a summary of our key findings and their implications in the final section.

\section{\label{sec:level2}THEORETICAL FORMALISM}
The equilibrium structure for a non-rotating, spherically symmetric NS with isotropic matter can be derived in the framework of General Relativity (GR) from the Tolman-Oppenheimer-Volkoff (TOV) \cite{Tolman1939,OppenheimerVolkoff1939} equations at hydrostatic equilibrium,
\begin{eqnarray}
	\label{eqn:TOV}
	\frac{dP(r)}{dr}&=&-\frac{G\,\varepsilon (r)\,M(r)}{{{c}^{2}}{{r}^{2}}}\left( 1+\frac{P(r)}{\varepsilon (r)} \right)\nonumber\\
	&\times&\left( 1+\frac{4\,\pi \,{{r}^{3}}P(r)}{{{c}^{2}}M(r)} \right){{\left( 1-\frac{2\,GM(r)}{r\,{{c}^{2}}} \right)}^{-1}}, \nonumber\\
	\frac{dM(r)}{dr}&=&\frac{4\,\pi \,{{r}^{2}}\varepsilon (r)}{{{c}^{2}}},
\end{eqnarray}
where $M(r)$ denotes the gravitational mass contained up to a radial distance ($r$) from the NS center, and $G$ is the gravitational constant. $\varepsilon(r)$ and $P(r)$ define the energy density and pressure profiles, respectively. The moment of inertia for slowly rotating NSs can be obtained, to first order, by solving Einstein's equations using the Hartle-Thorne metric expansion \cite{Hartle1967,HartleThorne1968}. Within this order of approximation, the MI is independent of the NS spin frequency \cite{HartleThorne1968,FriedmanStergioulas2013}. For slowly rotating NS, the angular velocity ($\Omega$) is assumed to be significantly smaller than its Keplerian limit (${\Omega}_{K}$) defining the maximum rotational frequency before the star would disintegrate, $\Omega\ll{\Omega}_{K}\sim \sqrt{GM/{R}^3}$. This allows the rotational effects to be considered as small perturbations to the static, spherically symmetric structure of the star \cite{Hartle1967,HartleThorne1968,FriedmanStergioulas2013,Glendenning2000}. Angular velocity above ${\Omega}_{K}$ would eject fluid content of NS into space. For slowly rotating, spherically symmetric NS, the perturbed line element ${ds}_{SR}^2$ can be obtained in terms of the non-rotating space time metric ${ds}_{0}^2$ as \cite{Hartle1967,HartleThorne1968,FriedmanStergioulas2013}
\begin{eqnarray}
	\label{eqn:dsSR2}
	ds_{SR}^{2}&=&ds_{0}^{2}-2\omega (r){{r}^{2}}{{\sin }^{2}}\theta \,dt\,d\phi\nonumber\\
	&=&-{{e}^{-2\nu (r)}}{{(dt)}^{2}}+{{e}^{2\lambda (r)}}{{(dr)}^{2}}+{{r}^{2}}{{(d\theta )}^{2}}\nonumber\\
	&+&{{r}^{2}}{{\sin }^{2}}\theta {{(d\phi )}^{2}}-2\omega (r){{r}^{2}}{{\sin }^{2}}\theta \,dt\,d\phi.  
\end{eqnarray}
The radially dependent metric functions, $\lambda(r)$ and $\nu(r)$, are derived in GR from the interior Schwarzschild solution \cite{Tolman1939,MisnerThorne1973,Hartle2003,Weinberg1972}, ${{e}^{2\lambda (r)}}={{\left( 1-\frac{2GM(r)}{r{{c}^{2}}} \right)}^{-1}}$, and ${{e}^{2\nu (r)}}={{e}^{-2\lambda (r)}}=\left( 1-\frac{2GM(r)}{r{{c}^{2}}} \right),\,\text{for}\,r\ge R$. For $r<R$, $\nu(r)$ is constrained by the condition ${{e}^{\nu (r)}}\mu (r)=\mu (R)\sqrt{1-\frac{2GM(R)}{R{{c}^{2}}}}$, where the chemical potential $\mu(r)$ is defined at a given baryonic density $\rho(r)$ as $\mu (r)=\frac{\varepsilon (r)+P(r)}{\rho (r)}$. The angular momentum contribution $j(r)$ from a sphere of radius $r$ within a rotating NS of mass $M(r)$ reads \cite{Hartle1967,HartleThorne1968,FriedmanStergioulas2013}
\begin{equation}
	\label{eqn:jor}
	j(r)={{e}^{-\nu (r)}}{{e}^{-\lambda (r)}}={{e}^{-\nu (r)}}\sqrt{1-\frac{2GM(r)}{r{{c}^{2}}}}.
\end{equation} 
This expression arises from the Hartle-Thorne formalism for slowly rotating relativistic stars. Using $\frac{dM}{dr}=\frac{4 \pi r^2 \varepsilon(r)}{c^2}$ we get,
\begin{eqnarray}
	\label{eqn:djdr}
	\frac{dj}{dr}&=&-{{e}^{-\nu (r)}}\frac{d\nu }{dr}\sqrt{1-\frac{2GM(r)}{r{{c}^{2}}}}\nonumber\\
	&-&\frac{G{{e}^{-\nu (r)}}\left( 4\pi {{r}^{3}}\varepsilon (r)-M(r){{c}^{2}} \right)}{{{r}^{2}}{{c}^{4}}\sqrt{1-\frac{2GM(r)}{r{{c}^{2}}}}}. 
\end{eqnarray}
Using
\begin{eqnarray}
	\label{eqn:dnudr}
	\frac{d\nu }{dr}&=&\frac{G}{{{c}^{2}}{{r}^{2}}}\left( M(r)+\frac{4\pi {{r}^{3}}P(r)}{{{c}^{2}}} \right){{\left( 1-\frac{2GM(r)}{r{{c}^{2}}} \right)}^{-1}}\nonumber\\
	&=&-\frac{dP(r)}{dr}{{\left( \varepsilon (r)+P(r) \right)}^{-1}}, 
\end{eqnarray}
we get,
\begin{equation}
	\label{eqn:djdrfinal}
	\frac{dj}{dr}=\frac{-4\pi rG{{e}^{-\nu (r)}}\left( \varepsilon (r)+P(r) \right)}{{{c}^{4}}\sqrt{1-\frac{2GM(r)}{r{{c}^{2}}}}},
\end{equation}
with total angular momentum ($J$) \cite{Hartle1967,HartleThorne1968,FriedmanStergioulas2013}
\begin{equation}
	\label{eqn:dJdrAng}
	\frac{dJ}{dr}=-\frac{2{{c}^{2}}}{3G}{{r}^{3}}\bar{\omega }\frac{dj}{dr}.
\end{equation}
Here, $\bar{\omega}=\Omega-\omega(r)$ is the local angular velocity at a certain point in NS relative to that of the local inertial frame. Using $J=\tilde{\kappa}\,\Omega$, where $\tilde{\kappa}$ is the MI at radial distance $r$ and $\Omega$ is the total angular velocity (frequency) of the star, we obtain the radial derivative of the MI as \cite{Hartle1967,HartleThorne1968,FriedmanStergioulas2013}
\begin{equation}
	\label{eqn:dkapdrfinal}
	\frac{d\tilde{\kappa }}{dr}=\frac{8\pi {{r}^{4}}{{e}^{-\nu (r)}}\left( \varepsilon (r)+P(r) \right)}{3{{c}^{2}}\sqrt{1-\frac{2GM(r)}{r{{c}^{2}}}}}\tilde{\omega},
\end{equation} 
where $\tilde{\omega}=\bar{\omega}/\Omega$ is the dimensionless relative frequency, and $\bar{\omega}$ satisfies \cite{Hartle1967,HartleThorne1968,FriedmanStergioulas2013,Glendenning2000,Tolman1939,MisnerThorne1973,Hartle2003,Weinberg1972}
\begin{equation}
	\label{eqn:DEOB}
	\frac{1}{{{r}^{4}}}\frac{d}{dr}\left[ {{r}^{4}}j\frac{d\bar{\omega }}{dr} \right]+\frac{4}{r}\frac{dj}{dr}\bar{\omega }=0.
\end{equation} 
We then get
\begin{equation}
	\label{eqn:dotdr}
	\frac{d\tilde{\omega }}{dr}=\frac{6G}{{{r}^{4}}{{c}^{2}}}{{e}^{\nu (r)}}{{\left( 1-\frac{2GM(r)}{r{{c}^{2}}} \right)}^{-1/2}}\tilde{\kappa }.
\end{equation}
Equations (\ref{eqn:dkapdrfinal}) and (\ref{eqn:dotdr}) can be solved simultaneously using the fourth order Runge-Kutta method, subject to the conditions $\bar{\omega}$ is regular at the center $\left({\left. \frac{d\tilde{\omega }}{dr} \right|}_{r=0}=0\right)$,
\begin{equation}
	\label{eqn:kappaoR}
	\tilde{\kappa}(R)=I=\frac{{{R}^{4}}{{c}^{2}}}{6G}{{\left. \frac{d\tilde{\omega }}{dr} \right|}_{r=R}},
\end{equation}
and
\begin{equation}
	\label{eqn:otoR}
	\tilde{\omega }(R)=1-\frac{2GI}{{{c}^{2}}{{R}^{3}}}.
\end{equation}
The total MI ($I$) of the star is then given by Eq. (\ref{eqn:kappaoR}). For slowly rotating NS, the total MI can be decomposed into contributions from the core (${I}_{\text{core}}$) and the crust (${I}_{\text{crust}}$), with the crustal fraction is then defined as \cite{HartleThorne1968,BejgerHaensel2002} ${{f}_{\text{crust}}}=\frac{\Delta I}{I}=\frac{{{I}_{\text{crust}}}}{I}$. The crustal fraction ($\Delta I \equiv {I}_{\text{crust}}$) is essential for comprehending the rotational dynamics, glitch phenomena, and thermal evolution of NSs \cite{HartleThorne1968,BejgerHaensel2002}. It can be obtained using the differential equation, 
\begin{equation}
	\label{eqn:dkcrustdr}
	\frac{d{{{\tilde{\kappa }}}_{cr}}}{dr}=\frac{8\pi {{r}^{4}}{{e}^{-\nu (r)}}\left( \varepsilon (r)+P(r) \right)}{3{{c}^{2}}\sqrt{1-\frac{2GM(r)}{r{{c}^{2}}}}}\tilde{\omega },
\end{equation}
which can be integrated from $r={R}_{\text{core}}$ (${\tilde{\kappa }}_{cr}=0$) up to $r=R$ (${\tilde{\kappa }}_{cr}={I}_{\text{crust}}$). Pulsar glitches, sudden spin-ups followed by gradual relaxation \cite{HaenselPotekhin2007}, arise from angular momentum transfer between the superfluid neutrons in the NS interior and its crust. The crust extends to the neutron drip density point where finite nuclei dissolve into homogeneous NM, with its fractional MI determining glitch magnitude. A larger crustal fraction stores more angular momentum, enabling stronger glitches.\par
For cold, uniform $npe\mu$ matter in $\beta$-equilibrium within the NS core, the total energy density $\varepsilon(r,\rho)$ and pressure are obtained by summing the baryonic ${\varepsilon}_{b}(p,n)$ and leptonic ${\varepsilon}_{e,\mu}$ contributions, in terms of the proton (${x}_{p}$), electron (${x}_{e}$), and muon (${x}_{\mu}$) fractions \cite{XuChen2009,Seif2011,SeifHashemJPG2024},
\begin{equation}
	\label{eqn:epstot}
	\varepsilon (\rho ,{{x}_{p}},{{x}_{e}},{{x}_{\mu }})={{\varepsilon }_{b}}(\rho ,{{x}_{p}})+\sum\limits_{\ell =e,\mu }{{{\varepsilon }_{\ell }}(\rho ,{{x}_{\ell }})},
\end{equation}
and
\begin{eqnarray}
	\label{eqn:Ptot}
	P(\rho ,{{x}_{p}},{{x}_{e}},{{x}_{\mu }})&=&{{P }_{b}}(\rho ,{{x}_{p}})+\sum\limits_{\ell =e,\mu }{{{P }_{\ell }}(\rho ,{{x}_{\ell }})} \nonumber\\
	&=&{{\rho }^{2}}\frac{\partial {({E}_{A}(\rho ,{{x}_{p}}))}}{\partial \rho} \nonumber\\
	&+&\sum\limits_{\ell =e,\mu }{\left[{{\mu }_{\ell }}\,\rho \,{{x}_{\ell }}-{{\varepsilon }_{\ell }}(\rho ,{{x}_{\ell }})\right]}.
\end{eqnarray} 
The total baryon energy reads
\begin{equation*}
	{{\varepsilon }_{b}}(\rho ,{{x}_{p}})=\rho \,\left[ {E}_{A}(\rho ,{{x}_{p}})+{{x}_{p}}\,{{m}_{p}}{{c}^{2}}+(1-{{x}_{p}})\,{{m}_{n}}{{c}^{2}} \right].
\end{equation*}
Using the effective CDM3Y density-dependent form \cite{KhoaSatchler1997} of the M3Y-Paris (Reid) \cite{AnantaramanToki1983,BertschBorysowicz1977} nucleon-nucleon (NN) interactions, the energy per nucleon (${E}_{A}$) can be expressed as \cite{Seif2011,SeifHashem2023},
\begin{eqnarray}
	\label{eqn:EDA}
	&&{E}_{A}\left(\rho ,{{x}_{p}}\right)=\frac{3\,{{\hbar }^{2}}\,k_{F}^{2}\left[ {{(2-2{{x}_{p}})}^{5/3}}+{{(2{{x}_{p}})}^{5/3}} \right]}{20\,m}\nonumber\\
	&&+\frac{\rho }{2}\,\left\{ {{F}_{0}}(\rho )\,J_{00}^{D}+{{(1-2{{x}_{p}})}^{2}}\,{{F}_{1}}(\rho )\,J_{01}^{D} \right\} \nonumber\\
	&&+\frac{\rho }{8}\int{d\vec{r}\left[ {{F}_{0}}(\rho )v_{00}^{Ex}{{({{B}_{0}})}^{2}}+{{F}_{1}}(\rho )v_{01}^{Ex}{{({{B}_{1}})}^{2}} \right]},
\end{eqnarray}
where ${{B}_{0}(r,{x}_{p})}=2\left[(1-{{x}_{p}})\,{{{\hat{J}}}_{1}}({{k}_{Fn}}r)+{{x}_{p}}\,{{{\hat{J}}}_{1}}({{k}_{Fp}}r)\right]$ and ${{B}_{1}(r,{x}_{p})}=2\left[(1-{{x}_{p}})\,{{{\hat{J}}}_{1}}({{k}_{Fn}}r)-{{x}_{p}}\,{{{\hat{J}}}_{1}}({{k}_{Fp}}r)\right]$. The first term in Eq. (\ref{eqn:EDA}) corresponds to the kinetic energy contribution, and $J_{00(01)}^{D}=\int{d\vec{r}\,v_{00(01)}^{D}(r)}$, where $v_{00(01)}^{D}(r)$ denote the central isoscalar (00) and isovector (01) components of the direct part of the M3Y interaction \cite{SatchlerLove1979}. The terms $v_{00(01)}^{Ex}(r)$ represent the corresponding exchange components. ${{\hat{J}}_{1}}(x)$ is defined using the first-order spherical Bessel function, ${{\hat{J}}_{1}}(x)=3{{j}_{1}}(x)/x$. The neutron and proton Fermi momenta are defined as ${{k}_{Fn(p)}}={{(3{{\pi }^{2}}{{\rho }_{n(p)}})}^{1/3}}$, while the total Fermi momentum is ${{k}_{F}}={{(3\,{{\pi }^{2}}\rho /2)}^{1/3}}$. To consider the medium effect, the isoscalar (${F}_{0}$) and isovector (${F}_{1}$) density dependencies have been incorporated into the bare M3Y interaction \cite{KhoaSatchler1997},
\begin{equation}
	\label{eqn:DensDep}
	\begin{aligned}
		& v_{00(01)}^{D,Ex}(\rho ,r)={{F}_{0(1)}}(\rho )\,v_{00(01)}^{D,Ex}(r), \\ 
		& {{F}_{0(1)}}(\rho )={{C}_{0(1)}}\left( 1+{\alpha}_{0(1)}\,{{e}^{-{\beta}_{0(1)}\,\rho }}-{\gamma}_{0(1)}\,\rho  \right). \\ 
	\end{aligned}
\end{equation}
Multiple parameterizations of the CDM3Y-${K}_{0}$ form (${C}_{i=0,1}$, ${\alpha}_{i}$, ${\beta}_{i}$ and ${\gamma}_{i}$) have been developed for isoscalar \cite{Seif2011} and isovector \cite{SeifHashem2023,SeifHashemJPG2024} density dependencies, allowing for a wide spectrum of EOS stiffness with isoscalar saturation incompressibility values ${K}_{0}=150-330$ MeV. The isoscalar part was optimized for symmetric nuclear matter (SNM) properties in a non-relativistic HF approach, while the isovector terms \cite{SeifHashem2023} were fitted \cite{KhoaThan2007} to match earlier BHF results for isovector neutron potential \cite{JeukenneLejeune1977,Lejeune1980}. The energy density of electrons and muons can be described using the non-interacting Fermi gas model \cite{Tolman1939,XuChen2009,SeifHashemJPG2024,OppenheimerVolkoff1939},
\begin{eqnarray}
	\label{eqn:epslep}
	&&{{\varepsilon }_{\ell =e(\mu )}}(\rho ,{{x}_{\ell }})=\frac{{{m}_{\ell }}\,{{c}^{2}}}{8\,{{\pi }^{2}}\,\lambda _{\ell }^{3}}[\lambda _{\ell }{{k}_{F\ell }}(\rho ,{{x}_{\ell }}) \nonumber\\
	&&\times \sqrt{1+{{({{\lambda }_{\ell }}\,{{k}_{F\ell }}(\rho ,{{x}_{\ell }}))}^{2}}} ( 1+2\,{{({{\lambda }_{\ell }}\,{{k}_{F\ell }}(\rho ,{{x}_{\ell }}))}^{2}} )  \nonumber\\
	&&-\ln ( {{\lambda }_{\ell }}\,{{k}_{F\ell }}(\rho ,{{x}_{\ell }})+\sqrt{1+{{({{\lambda }_{\ell }}\,{{k}_{F\ell }}(\rho ,{{x}_{\ell }}))}^{2}}} )],
\end{eqnarray}
where ${{\lambda }_{\ell }}=\hbar c/{{m}_{\ell }}{{c}^{2}}$ and ${{k}_{F\ell}}={{(3{{\pi }^{2}}{{\rho }_{\ell}})}^{1/3}}$ denote the leptonic Compton wavelength and Fermi momentum, respectively. The corresponding chemical potential can be expressed in terms of the rest mass energy (${{m}_{\ell }}{{c}^{2}}$) and ${k}_{F\ell}$ as,
\begin{equation}
	\label{eqn:mul}
	{{\mu }_{\ell }}(\rho ,{{x}_{\ell }})=\sqrt{{{\hbar }^{2}}{{c}^{2}}{{({{k}_{F\ell }}(\rho ,{{x}_{\ell }}))}^{2}}+m_{\ell }^{2}{{c}^{4}}}.
\end{equation} 
The DU process operates efficiently when the Fermi momenta of participating particles satisfy kinematic conditions for energy and momentum conservation. In contrast to the slower modified Urca process, DU activation depends critically on specific thresholds in density and proton fraction. Based on the present CDM3Y EOSs, we can determine such threshold values for DU in $npe\mu$ NS core matter. In $\beta$-stable $npe\mu$ matter, the DU reactions ($n\rightarrow p+{e}^{-}+{\tilde{\nu}}_{e}$ and $p+{e}^{-}\rightarrow n+{\nu}_{e}$) maintain chemical equilibrium, thermodynamically implying \cite{SeifHashem2022,LoanTan2011,SteinerPrakash2005,SeifBasu2014}
\begin{equation}
	\label{eqn:equill}
	{{\mu }_{e}}={{\mu }_{\mu }}={{\mu }_{n}}-{{\mu }_{p}}=-\frac{\partial{({E}_{A}(\rho ,{{x}_{p}}))}}{\partial {{x}_{p}}}.
\end{equation}
At $T = 0$ MeV, the charge neutrality condition (${x}_{p}={x}_{e}+{x}_{\mu}$) together with Eqs. (\ref{eqn:mul}) and (\ref{eqn:equill}) allow calculation of charge-particle fractions by solving the resulting equations \cite{LoanTan2011,SeifHashem2022},
\begin{equation}
	\label{eqn:pemufrac}
	3{{\pi }^{2}}{{(\hbar c)}^{3}}\rho {{x}_{p}}-\mu _{e}^{3}-{{\left[ \mu _{e}^{2}-{{({{m}_{\mu }}{{c}^{2}})}^{2}} \right]}^{3/2}}\Theta ( {{\mu }_{e}}-{{m}_{\mu }}{{c}^{2}})=0.
\end{equation}
The Heaviside step function $\Theta({{\mu }_{e}}-{{m}_{\mu }}{{c}^{2}})$ is zero for ${\mu }_{e}<m_\mu c^2$ and unity otherwise. The DU process becomes allowed when the proton fraction exceeds a threshold value expressible in terms of the leptonic electron-muon fraction ${r}_{e}=1/\left[1+({x}_{\mu}/{x}_{e})\right]$ \cite{SeifHashem2022,LoanTan2011,KlahnBlaschke2006},
\begin{equation}
	\label{eqn:xDU}
	{{x}_{\text{DU}}}={{x}_{p}}(\text{DU})=\frac{1}{1+{{\left( 1+r_{e}^{1/3} \right)}^{3}}}.
\end{equation}
The threshold DU proton fraction is then ${x}_{\text{DU}}=1/9$ when ${\mu }_{e}<m_\mu c^2$, otherwise ${x}_{\text{DU}}>1/9$ \cite{SeifHashem2022,LoanTan2011,LattimerPethick1991,LattimervanRiper1994,Lattimer2018}. \par
By requiring the compressibility (${K}_{{{\mu }_{i}}}$) at a specific chemical potential (${\mu}_{i}$) to remain positive, one can determine the transition density (${\rho}_{t}$) and its corresponding pressure (${P}_{t}$), and proton fraction (${x}_{pt}$) at the interface between the NS solid crust and its liquid NS core \cite{SeifBasu2014,Kubis2007},
\begin{equation}
	\label{eqn:kmu}
	{{K}_{{{\mu }_{i}}}}={{\left( \frac{\partial P}{\partial \rho } \right)}_{{{\mu }_{i}}}}=\frac{K(\rho ,{x}_{p})}{9}-\frac{{{\left( \rho\, \frac{{{\partial }^{2}}{({E}_{A})}}{\partial \rho \partial {x}_{p}} \right)}^{2}}}{\frac{{{\partial }^{2}}{({E}_{A})}}{\partial {{{x}_{p}}^{2}}}}>0.
\end{equation}
The second term in this equation arises from the contribution of lepton pressure. $K(\rho,{x}_{p})$ is the incompressibility coefficient of ANM at a specific proton fraction (${x}_{p}$) \cite{Seif2011,SeifHashemJPG2024,SeifHashem2023}. \par
The adiabatic index ($\varGamma$) is essential for understanding the mechanical and thermodynamic properties of NS, as it quantifies the pressure-density relation in adiabatic processes. It is fundamentally linked to the stability of compact stars against radial oscillations. In the high-density NS interior, $\varGamma$ becomes crucial for stability analysis, equilibrium structure, and dynamic evolution. It is formally defined as the logarithmic derivative of pressure with respect to mass density under adiabatic conditions, and it can be expressed in terms of $P$, $\varepsilon$, and speed of sound (${v}_{s}$) as
\begin{equation}
	\label{eqn:Adiaindex}
	\varGamma =\frac{\varepsilon +P}{P}\left( \frac{dP}{d\varepsilon } \right)=\left( \frac{\varepsilon +P}{P} \right){{\left( \frac{{{v}_{s}}}{c} \right)}^{2}}.
\end{equation}
NS stability requires the volume-averaged adiabatic index to be greater than 4/3. The adiabatic index asymptotically approaches infinity in the limit of perfectly incompressible matter with maximal stiffness.  Also, the NS compactness can be estimated by comparing its gravitational Schwarzschild radius, its critical size at which it would become a black hole ($2GM/c^2$), to its actual physical radius $R$, $\mathpzc{C}=\frac{G M}{R c^2}$. \par
Based on the charge symmetry of nuclear force, the nucleonic energy ${E}_{A} (\rho,\mathcal{I})$  in ANM can be expanded around SNM saturation density (${\rho}_{0}$) and isospin symmetry ($\mathcal{I}=0$) as,
\begin{widetext}
	\begin{equation}
		\label{eqn:EATEI}
		{{E}_{A}}(\rho,\mathcal{I})=\sum\limits_{n=0,1,2,...}{\frac{1}{(2n)!}}\frac{{{\partial }^{(2n)}}}{\partial {{\mathcal{I}}^{(2n)}}}{{\left. \left[ \sum\limits_{m=0,1,2,...}{{{\left. \frac{1}{m!}\frac{{{\partial }^{m}}{{E}_{A}}(\rho ,\mathcal{I})}{\partial {{\rho }^{m}}} \right|}_{{{\rho }_{0}}}}{{(\rho -{{\rho }_{0}})}^{m}}} \right] \right|}_{\mathcal{I}=0}}{{\mathcal{I}}^{(2n)}}.
	\end{equation}  
\end{widetext}
The functional dependence of ${{E}_{A}}(\rho,\mathcal{I})$ isospin asymmetry ($\mathcal{I}$) can be rigorously expressed as
\begin{widetext}
	\begin{eqnarray}
		\label{eqn:EAexpandrhoI}
		{E}_{A}(\rho ,\mathcal{I})&=&\sum\limits_{n=0,2,4,6,...}{\left[ {{E}_{symn}}({{\rho }_{0}})+{{L}_{n}}\mathbb{X}+\frac{{{K}_{n}}}{2!}{{\mathbb{X}}^{2}}+\frac{{{Q}_{n}}}{3!}{{\mathbb{X}}^{3}}+\frac{{{I}_{n}}}{4!}{{\mathbb{X}}^{4}}+\frac{{{H}_{n}}}{5!}{{\mathbb{X}}^{5}}+\frac{{{G}_{n}}}{6!}{{\mathbb{X}}^{6}}+...  \right]}\,{{\mathcal{I}}^{n}} \nonumber \\ 
		&\equiv& {{E}_{A}}(\rho ,0)+{{E}_{sym2}}(\rho ){{\mathcal{I}}^{2}}+{{E}_{sym4}}(\rho ){{\mathcal{I}}^{4}}+{{E}_{sym6}}(\rho){{\mathcal{I}}^{6}}+...,   
	\end{eqnarray} 
\end{widetext}
with ${{E}_{symn}}={{\left. \frac{1}{n!}\frac{{{\partial }^{n}}{{E}_{A}}(\rho ,\mathcal{I})}{\partial {{\mathcal{I}}^{n}}} \right|}_{\mathcal{I}=0}}$ and ${E}_{A}(\rho,0)\equiv{E}_{sym0}$. The energy density of SNM, ${E}_{A}(\rho,\mathcal{I}=0)$, along with the quadratic and quartic symmetry energy coefficients, ${E}_{sym2,4}(\rho)$, can be formulated as functions of the scaled density parameter $\mathbb{X}=\left( \frac{\rho -{{\rho }_{0}}}{3{{\rho }_{0}}} \right)$ as
\begin{widetext}
	\begin{eqnarray}
		\label{eqn:Esymexpandrho}
		{{E}_{A}}(\rho ,\mathcal{I}=0)&=&{{E}_{A}}({{\rho }_{0}},\mathcal{I}=0)+\frac{{{K}_{0}}}{2!}{{\mathbb{X}}^{2}}+\frac{{{Q}_{0}}}{3!}{{\mathbb{X}}^{3}}+\frac{{{\mathcal{I}}_{0}}}{4!}{{\mathbb{X}}^{4}}+\frac{{{H}_{0}}}{5!}{{\mathbb{X}}^{5}}+\frac{{{G}_{0}}}{6!}{{\mathbb{X}}^{6}}+..., \nonumber\\ 
		{{E}_{sym2}}(\rho )&\equiv& \frac{1}{2!}{{\left. \frac{{{\partial }^{2}}{{E}_{A}}(\rho ,\mathcal{I})}{\partial {{\mathcal{I}}^{2}}} \right|}_{\mathcal{I}=0}}={{E}_{sym}}({{\rho }_{0}})+L\mathbb{X}+\frac{{{K}_{2}}}{2!}{{\mathbb{X}}^{2}}+\frac{{{Q}_{2}}}{3!}{{\mathbb{X}}^{3}}+\frac{{{\mathcal{I}}_{2}}}{4!}{{\mathbb{X}}^{4}}+\frac{{{H}_{2}}}{5!}{{\mathbb{X}}^{5}}+\frac{{{G}_{2}}}{6!}{{\mathbb{X}}^{6}}+..., \nonumber\\ 
		{{E}_{sym4}}(\rho )&\equiv& \frac{1}{4!}{{\left. \frac{{{\partial }^{4}}{{E}_{A}}(\rho ,\mathcal{I})}{\partial {{\mathcal{I}}^{4}}} \right|}_{\mathcal{I}=0}}={{E}_{sym4}}({{\rho }_{0}})+{{L}_{4}}\mathbb{X}+\frac{{{K}_{4}}}{2!}{{\mathbb{X}}^{2}}+\frac{{{Q}_{4}}}{3!}{{\mathbb{X}}^{3}}+\frac{{{\mathcal{I}}_{4}}}{4!}{{\mathbb{X}}^{4}}+\frac{{{H}_{4}}}{5!}{{\mathbb{X}}^{5}}+\frac{{{G}_{4}}}{6!}{{\mathbb{X}}^{6}}+....
	\end{eqnarray}  
\end{widetext}
Now, the density dependence of ANM is quantified by the symmetry energy coefficients ${{L}_{n}}=3{{\rho }_{0}}{{\left. \frac{d{{E}_{symn}}}{d\rho } \right|}_{{{\rho }_{0}}}}$ with ($L={L}_{2}$), ${{K}_{n}}={{(3{{\rho }_{0}})}^{2}}{{\left. \frac{{{d}^{2}}{{E}_{symn}}}{d{{\rho }^{2}}} \right|}_{{{\rho }_{0}}}}$, ${{Q}_{n}}={{(3{{\rho }_{0}})}^{3}}{{\left. \frac{{{d}^{3}}{{E}_{symn}}}{d{{\rho }^{3}}} \right|}_{{{\rho }_{0}}}}$, ${{I}_{n}}={{(3{{\rho }_{0}})}^{4}}{{\left. \frac{{{d}^{4}}{{E}_{symn}}}{d{{\rho }^{4}}} \right|}_{{{\rho }_{0}}}}$, ${{H}_{n}}={{(3{{\rho }_{0}})}^{5}}{{\left. \frac{{{d}^{5}}{{E}_{symn}}}{d{{\rho }^{5}}} \right|}_{{{\rho }_{0}}}}$, and ${{G}_{n}}={{(3{{\rho }_{0}})}^{6}}{{\left. \frac{{{d}^{6}}{{E}_{symn}}}{d{{\rho }^{6}}} \right|}_{{{\rho }_{0}}}}$. Alternative notations ${K}_{sym}={K}_{2}$ and ${K}_{symn}={K}_{n}$ are sometimes used. Similarly, the isospin-asymmetry dependence of the ANM incompressibility can be expressed as \cite{Seif2012,HashemHassanien2024},
\begin{eqnarray}
	\label{eqn:K0I}
	{{K}_{0\mathcal{I}}}&\equiv& {{(3{{\rho }_{0\mathcal{I}}})}^{2}}{{\left. \frac{{{d}^{2}}{{E}_{A}(\rho,\mathcal{I})}}{d{{\rho }^{2}}} \right|}_{{{\rho }_{0I}}}}={{K}_{0}}({{\rho }_{0}},\mathcal{I}=0) \nonumber\\
	&+&{{K}_{\tau2}}{{\mathcal{I}}^{2}}+{K}_{\tau4}\,{\mathcal{I}}^{4}+{K}_{\tau6}\,{\mathcal{I}}^{6}+....
\end{eqnarray} 
The different CDM3Y-based symmetry energy coefficients are analytically derived explicitly in Ref. \cite{SeifHashem2025}, providing a complete set of isoscalar/isovector parameters for describing ANM's bulk and surface properties.

\section{\label{sec:level3}RESULTS AND DISCUSSION}
Beyond their established role in determining neutron star properties, including the mass, radius, compactness, tidal deformability, the speed of sound through its core, and surface characteristics \cite{SeifHashem2025}, variations in the density-isospin derivatives, from slope, curvature and skewness to kurtosis and the higher order isoscalar/isovector symmetry-energy coefficients in the nuclear EOS, are anticipated to impact the crust’s inner edge and moment of inertia. In this section, we systematically investigate how these EOS parameters affect the core-crust transition properties, the moment of inertia, and the crustal fraction contribution to the total NS moment of inertia. We examine a broad set of reliable CDM3Y-${K}_{0}$ equations of state based on the M3Y-Paris NN interaction, spanning SNM incompressibility (${K}_{0}$) over a broad range from soft (${K}_{0}$ = 150 MeV) to extremely stiff (330 MeV) NM. Derived within a non-relativistic HF framework, these reliable EOSs have been successfully applied to study both cold and hot nuclear matter, and to obtain the detailed structure of non-rotating NSs and their deformability.
\subsection{\label{sec:level3a} Core-crust transition density and pressure}
\begin{figure*}[!htbp] 
	\centering	
	\subfigure[ \label{fig:Fig1a}]{
		\includegraphics[height=12cm,width=1.0\linewidth]{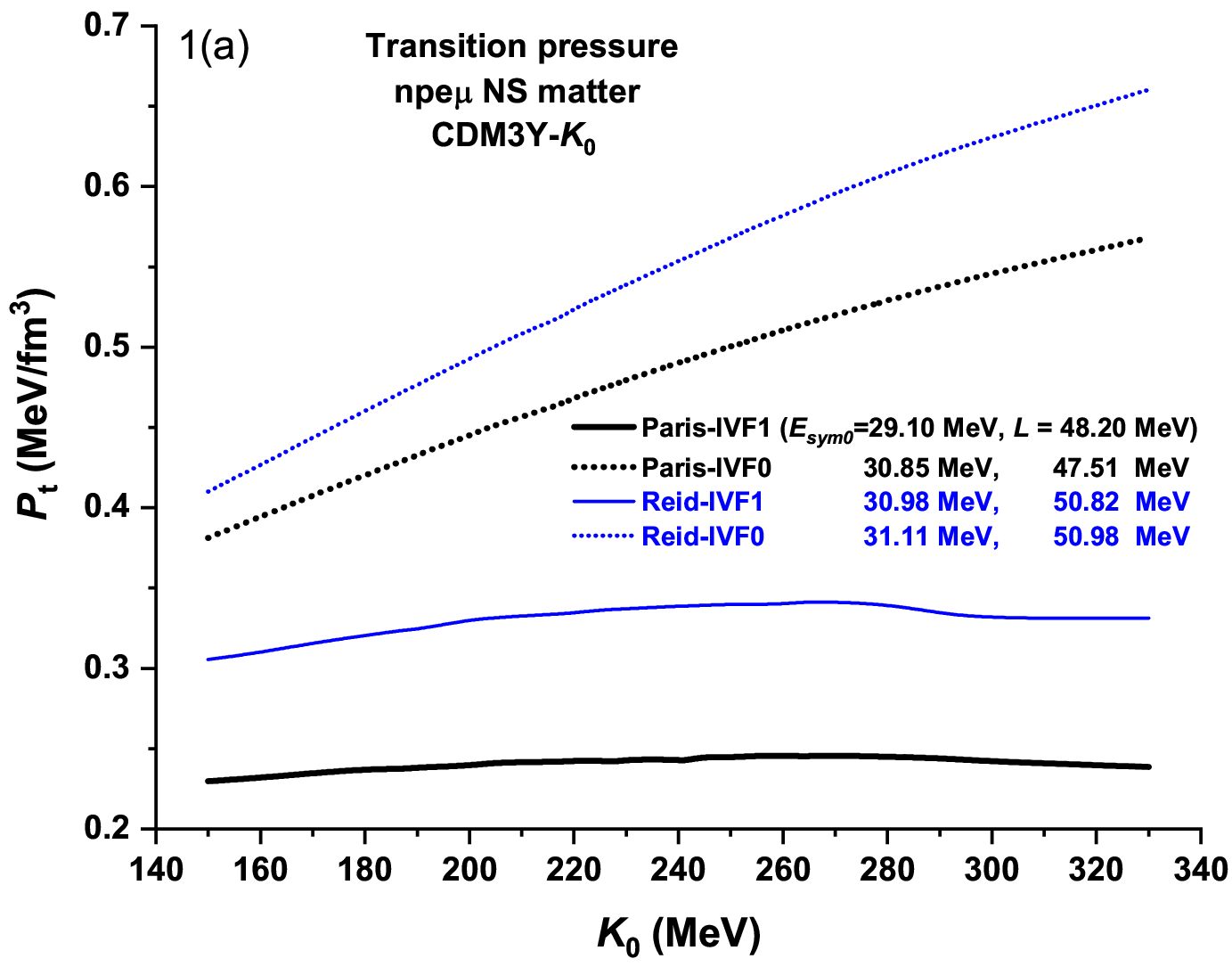}}
	~
	\centering
	\subfigure[ \label{fig:Fig1b}]{
		\includegraphics[height=12cm,width=1.0\linewidth]{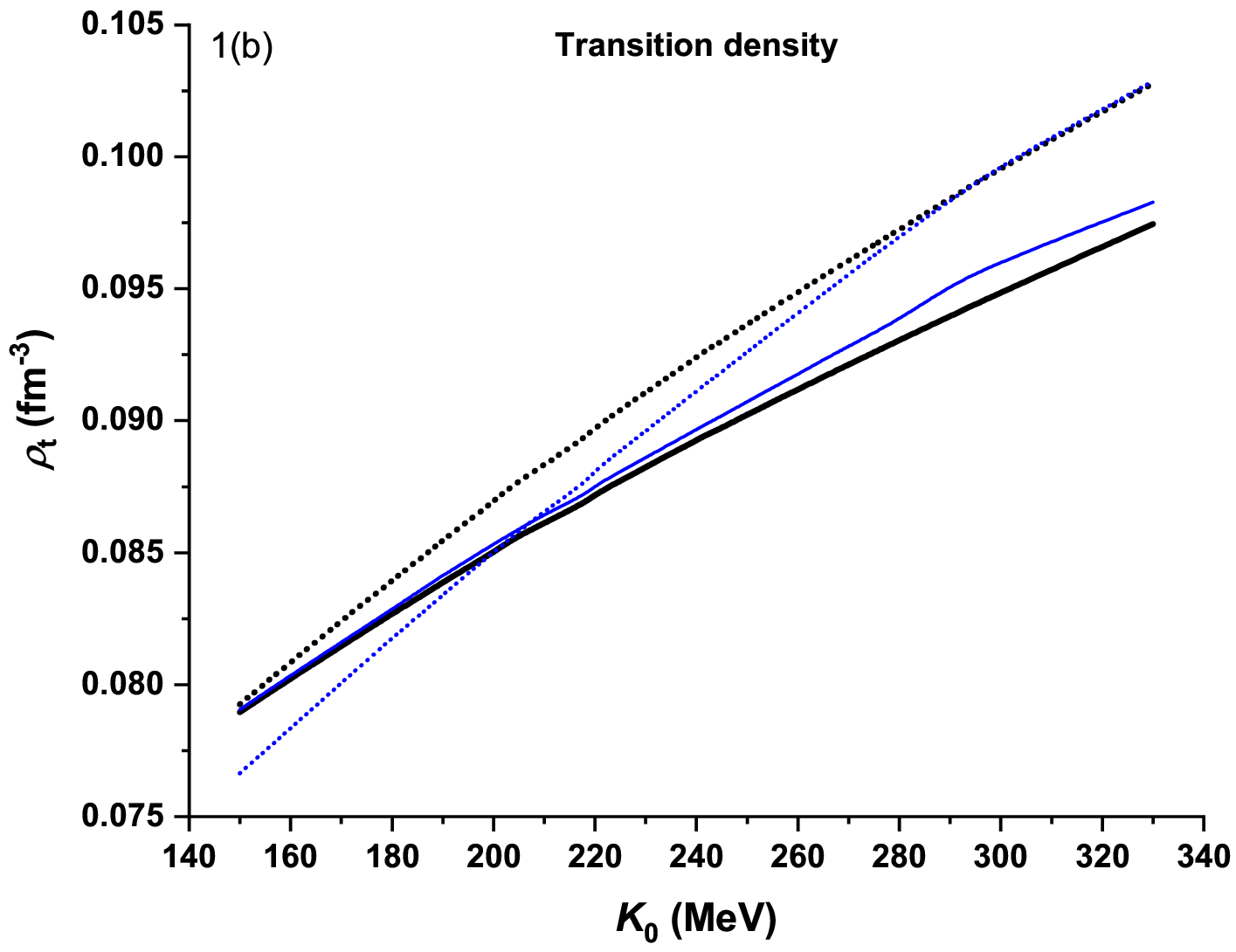}}	
\end{figure*}

\begin{figure*}[!htbp] 
	\centering	
	\subfigure[ \label{fig:Fig1c}]{
		\includegraphics[height=12cm,width=1.0\linewidth]{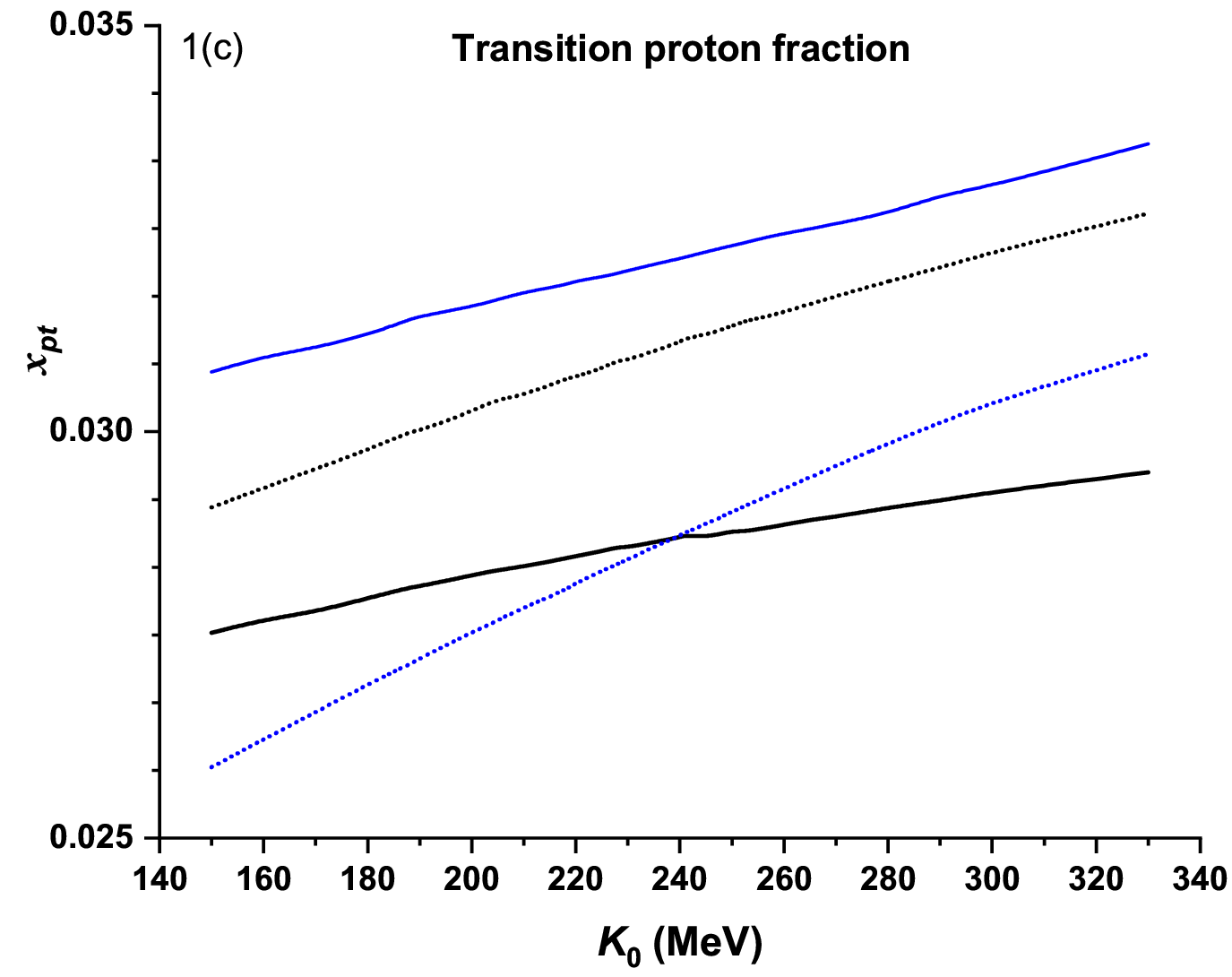}}
	~
	\centering
	\subfigure[ \label{fig:Fig1d}]{
		\includegraphics[height=12cm,width=1.0\linewidth]{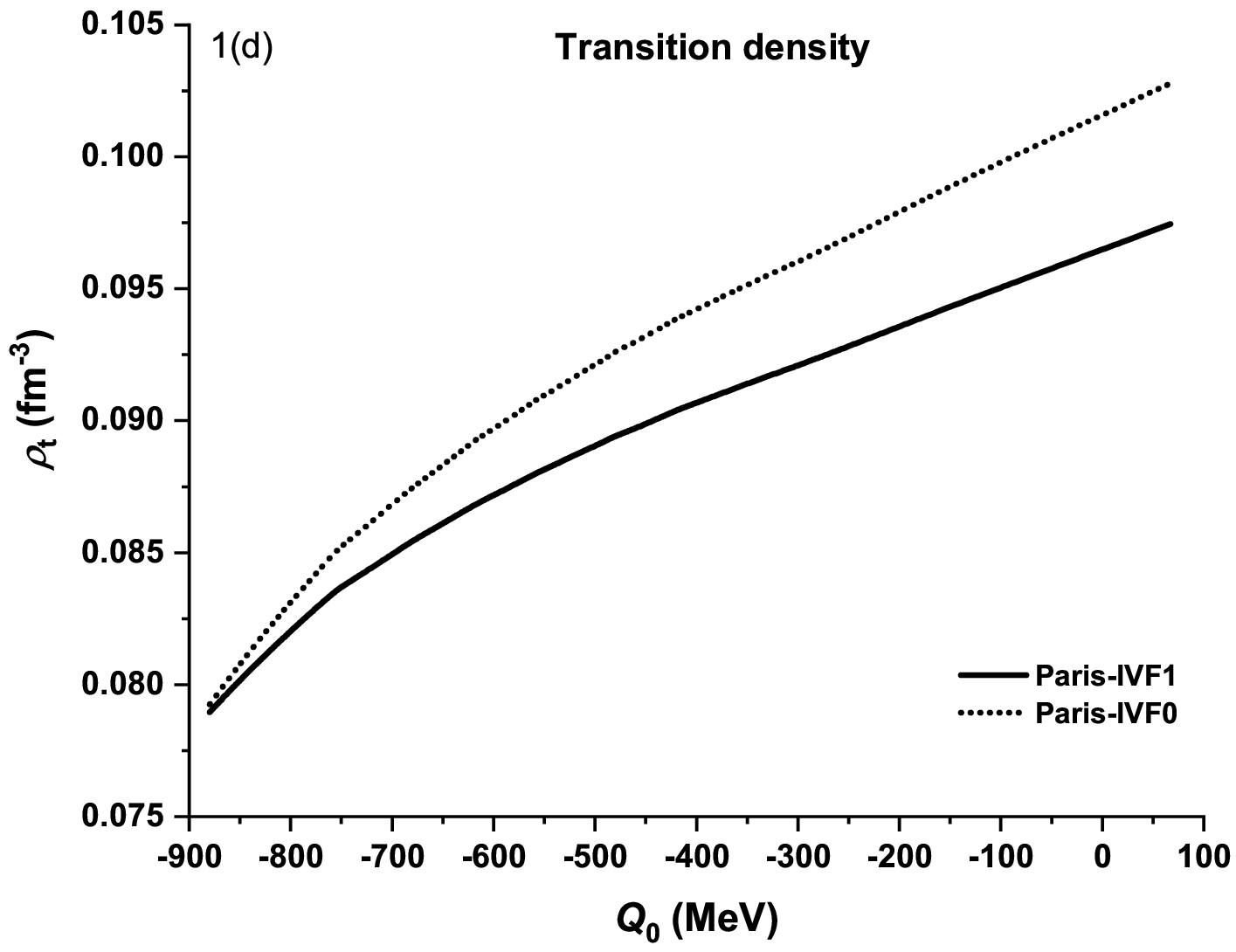}}	
\end{figure*}

Our initial investigation focuses on the influence of the stiffness of isospin-asymmetric $npe\mu$ matter and its associated EOS, across the inner and outer core, on the liquid core-solid crust transition density and pressure. The core-crust transition pressure (${P}_{t}$), derived from the incompressibility condition (Eq. (\ref{eqn:kmu})) using the EOSs based on the CDM3Y-Paris-IVF1 (${E}_{sym0}$ = 29.10 MeV, $L$ = 48.20 MeV), CDM3Y-Paris-IVF0 (30.85 MeV, 47.51 MeV), CDM3Y-Reid-IVF1 (30.98 MeV, 50.82 MeV), and CDM3Y-Reid-IVF0 (31.11 MeV, 50.98 MeV) NN interactions, is plotted in Fig. \ref{fig:Fig1a} against ${K}_{0}$. The associated transition density (${\rho}_{t}$) and proton fraction (${x}_{pt}$) are displayed in Figs. \ref{fig:Fig1b} and \ref{fig:Fig1c}, respectively. The isovector density dependence in the IVF1 versions of the M3Y-Paris and Reid interactions is derived independently of the isoscalar parameterization \cite{SeifHashemJPG2024,SeifHashem2023}, by comparing with BHF results for the nucleon optical potential \cite{JeukenneLejeune1977,Lejeune1980}. This parameterization predicts hard symmetry energy at high densities \cite{SeifHashem2023}. In contrast, the IVF0 isovector parameterization scales with the isoscalar density dependence to match experimental NM observables \cite{Seif2011} and charge-exchange reaction data \cite{KhoaThan2005}, resulting in a soft symmetry energy at supra-saturation densities \cite{Seif2011}. From Fig. \ref{fig:Fig1a}, we observe that the transition pressure increases with increasing the isoscalar symmetry energy coefficient ${E}_{sym0}$. Given nearly similar values of ${E}_{sym0}$(Paris-IVF0 and Reid-IVF1), ${P}_{t}$ diminishes with rising $L$. Figures \ref{fig:Fig1b} and \ref{fig:Fig1c} show that the transition density and proton fraction exhibit minimal variation across the differences in the symmetry energy coefficient ${E}_{sym0}$ and its slope $L$, with changes constrained to a narrow range of $0.003-0.007$. Both ${\rho}_{t}$ and ${x}_{pt}$ increase slightly with ${E}_{sym0}$. ${\rho}_{t}$ (${x}_{pt}$) slightly decreases (increases) with $L$ for nearly identical ${E}_{sym0}$ values. Generally, the core-crust transition pressure and density increase with increasing $L$ that increases alongside ${E}_{sym0}$. As nuclear matter stiffens, all three transition quantities increase in magnitude, though to differing degrees. This stiffness dependence weakens progressively from ${\rho}_{t}$ and ${P}_{t}$ to ${x}_{pt}$, which shows minimal sensitivity to nuclear incompressibility. This stiffness dependence also increases with increasing both ${E}_{sym0}$ and $L$. The influence of higher-order symmetry energy coefficients ${E}_{sym2,4,6,8}$ and the corresponding density slopes (${L}_{2,4,6,8}$) on core-crust transition properties is thoroughly investigated, by one of us, in Ref. \cite{SeifBasu2014}. Increasing the core-crust transition pressure, driven by a stiffer core NM EOS, indicates larger radius and greater tidal deformability of NS of a fixed mass \cite{SeifHashem2025,ArbanilRodrigues2023}. A higher transition proton fraction at larger densities facilitates enhanced neutrino emission, which in turn accelerates the neutron star cooling rate. \par
\begin{figure*}[!htbp]
	\subfigure[ \label{fig:Fig1e}]{
		\includegraphics[height=13cm,width=1.0\linewidth]{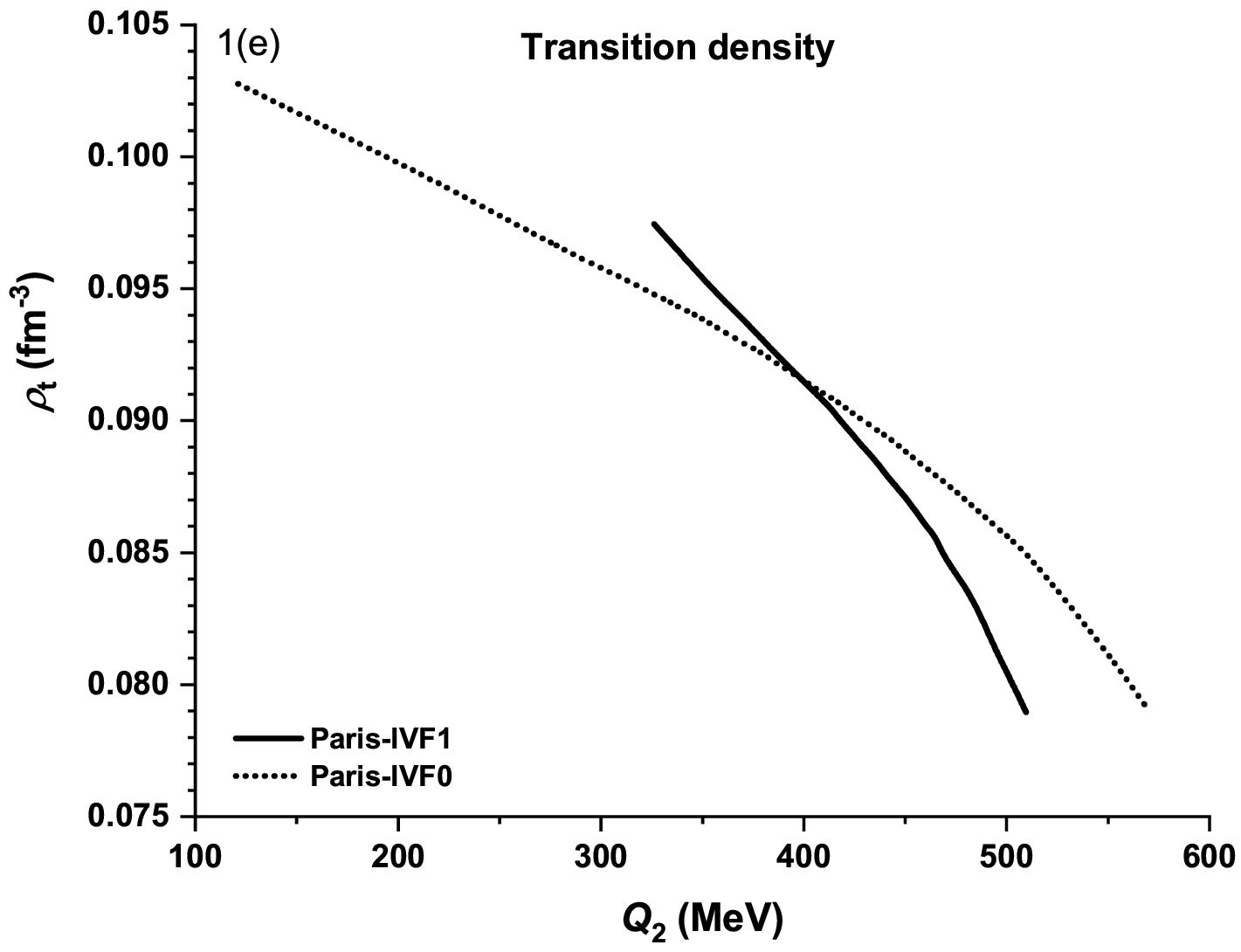}}
	\caption{\label{fig:Figure1} Core-crust transition (a) pressure, (b) density, and (c) proton fraction, using Eq. (\ref{eqn:kmu}), based on the CDM3Y-Paris-IVF1, CDM3Y-Paris-IVF0, CDM3Y-Reid-IVF1, and CDM3Y-Reid-IVF0 EOSs, as functions of ${K}_{0}$, and the dependence of the transition density on (d) the isoscalar skewness coefficient ${Q}_{0}$ and on (e) its isovector coefficient (${Q}_{2}$), based on CDM3Y-Paris-IVF1 and IVF0 EOSs.}  
\end{figure*}

Our analysis in Ref. \cite{SeifHashem2025} demonstrated that the stiffness of NM increases with increasing the fourth-order skewness coefficient (${Q}_{4}$) of the symmetry energy and with reducing the negativity of isoscalar coefficients ${Q}_{0}$ and ${H}_{0}$, and that of the isovector ${K}_{4}$, ${I}_{2,4}$, and ${G}_{2}$ coefficients, as well as that of the isobaric ${K}_{\tau4}$ symmetry coefficient. Then, increasing the positivity of these symmetry coefficients increases the core-crust transition density and pressure, and slightly the corresponding proton-fraction. As an example, Fig. \ref{fig:Fig1d} shows the transition density as a function of the isoscalar skewness coefficient ${Q}_{0}$. Consistent with expectations, ${\rho}_{t}$ increases when the negativity of ${Q}_{0}$ decreases. Conversely, increasing the stiffness of NM has been found to reduce the isoscalar symmetry coefficients ${I}_{0}$ and ${G}_{0}$, the isovector coefficients ${Q}_{2}$, ${K}_{6}$, and ${H}_{2}$ , as well as the isobaric coefficient ${K}_{\tau6}$ \cite{SeifHashem2025}. It also increases the negativity of ${K}_{2}$ and ${K}_{\tau2}$. As a result, the dependence of transition properties on these coefficients is expected to reverse their trends with ${K}_{0}$. For instance, larger values of ${I}_{0}$, ${G}_{0}$, ${Q}_{2}$, ${H}_{2}$, ${K}_{6}$ and ${K}_{\tau6}$ or a reduction in the negativity of ${K}_{2}$ and ${K}_{\tau2}$ should correspond to higher ${\rho}_{t}$ and ${P}_{t}$ , along with a modest increase in ${x}_{pt}$. To confirm, Fig. \ref{fig:Fig1e} displays the dependence of the transition density on the isovector skewness coefficient ${Q}_{2}$. Compared to IVF0, the isovector IVF1 density dependence restricts ${Q}_{2}$ to a narrower range, for the same range of ${K}_{0}=150-330$ MeV. Consistent with predictions, ${\rho}_{t}$ exhibits a decreasing trend with ${Q}_{2}$. Investigations based on Negele-Vautherin EOSs \cite{ZhangLi2018,ZhangLi2025} confirm that higher ${K}_{2}$ negativity correlates with reduced core-crust transition density and pressure. Furthermore, analyses using 21 EOSs derived from the Lagrangian description of NM \cite{MalikPais2024}, combined with compressible liquid-drop approximations for the inner crust and with corresponding dynamical spinodal calculations, indicated the increasing behavior of ${\rho}_{t}$ and ${P}_{t}$ with the stiffness of ANM, and a decreasing trend with ${I}_{0}$. The increasing trend of ${\rho}_{t}$ with ${K}_{2}$ and ${K}_{\tau2}$ can be confirmed also from the calculations based on six families of relativistic mean field models, incorporating nonlinear terms up to fourth order and dynamical spinodals \cite{PaisProvidencia2016}.
\subsection{\label{sec:level3b} Central particle-contributions} 
\begin{figure*}[!htbp] 
	\centering	
	\subfigure[ \label{fig:Fig2a}]{
		\includegraphics[height=12cm,width=1.0\linewidth]{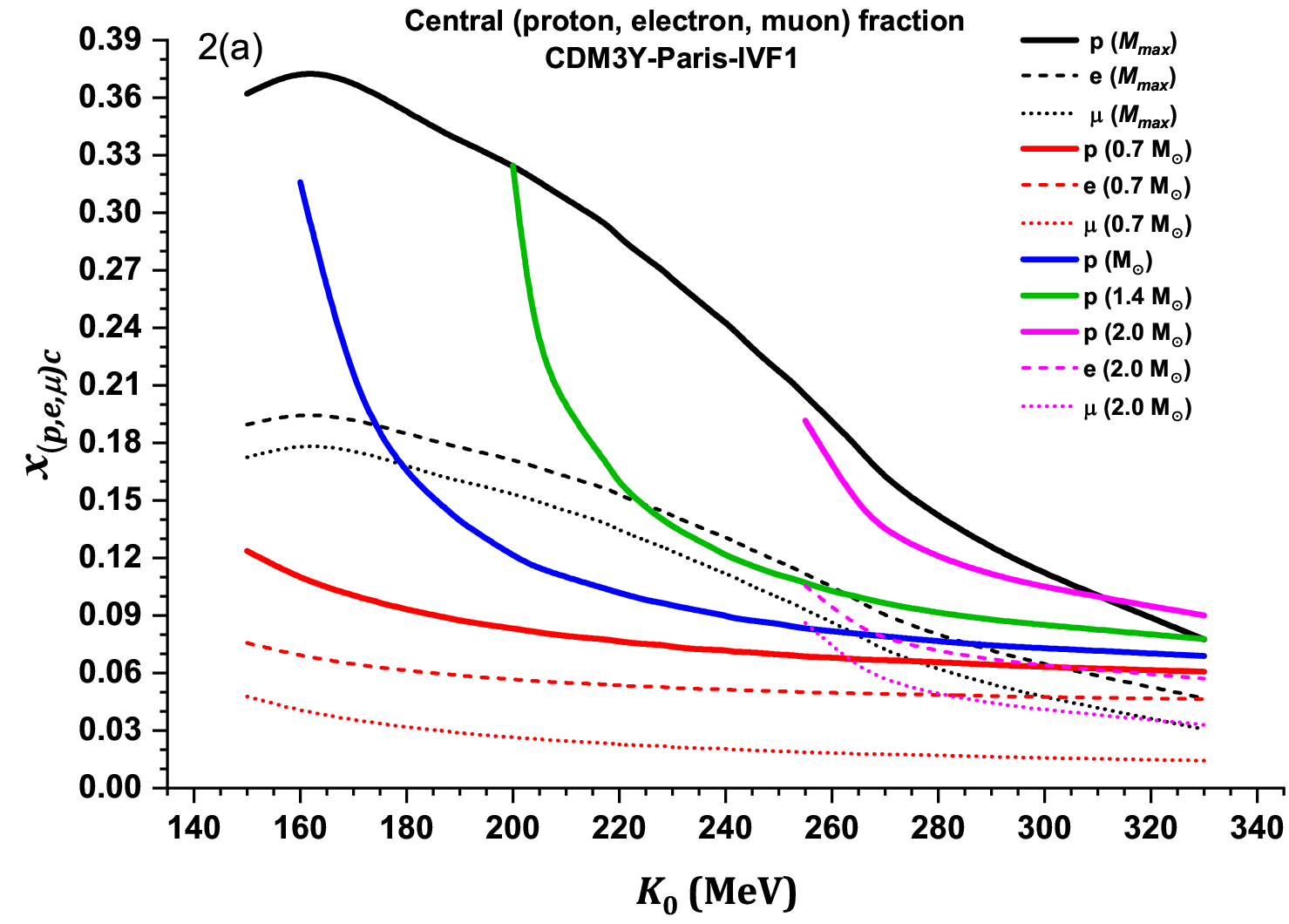}}
	~
	\centering
	\subfigure[ \label{fig:Fig2b}]{
		\includegraphics[height=12cm,width=1.0\linewidth]{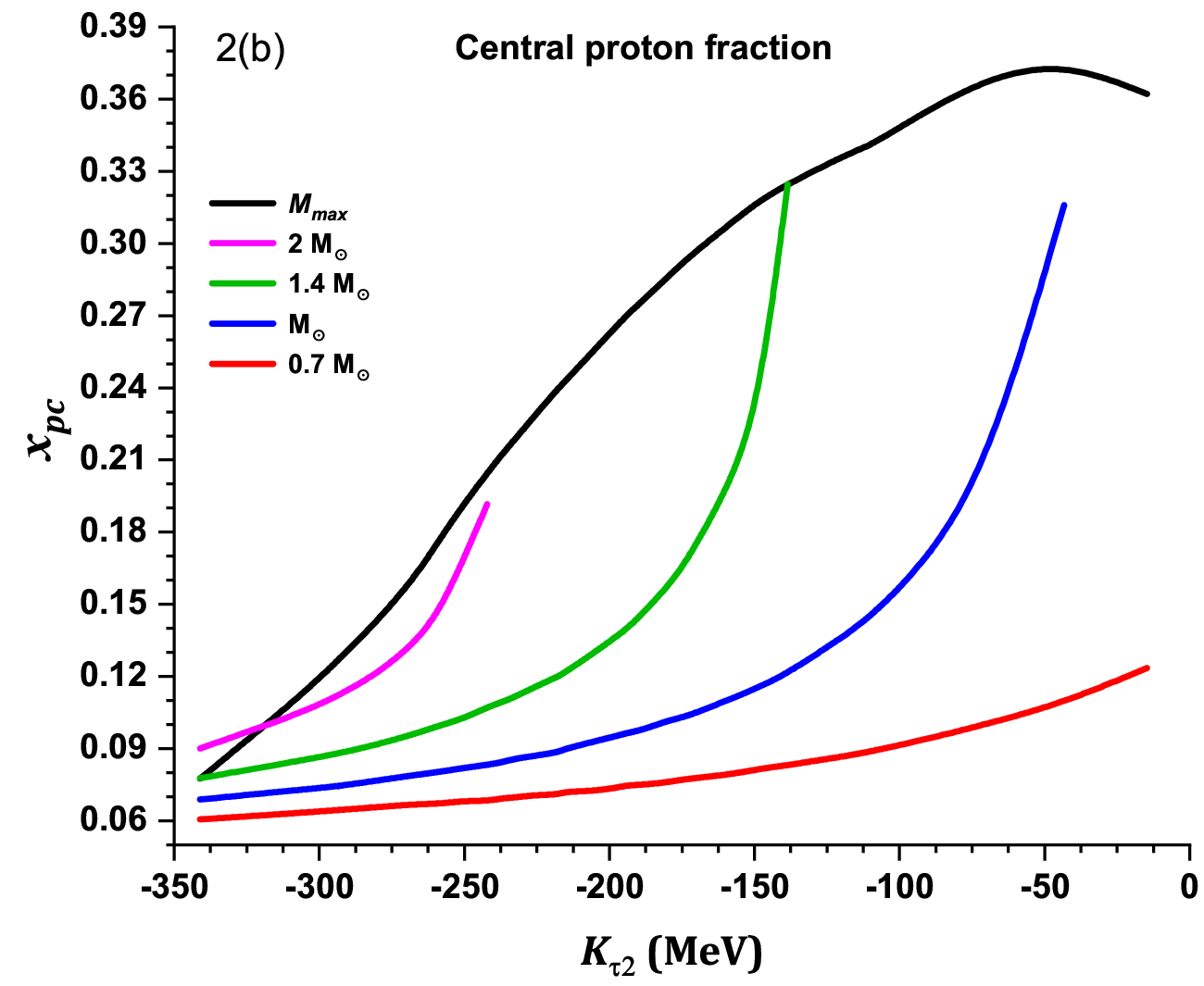}}	
\end{figure*}

The fractional particle contributions determine the isospin-asymmetry of NS core matter, which in turn governs how symmetry energy influences most NS properties. This asymmetry, along with the resulting particle fractions, is determined by the stiffness of the NS core matter, most critically within the dense inner core. Figure \ref{fig:Fig2a} displays the central proton (${x}_{pc}$), electron (${x}_{ec}$), and muon (${x}_{\mu c}$) fractions in cold $\beta$-equilibrated $npe\mu$ NS matter, across stellar masses from 0.7 ${\text{M}}_{\odot}$ to 2 ${\text{M}}_{\odot}$, as functions of the incompressibility ${K}_{0}$ coefficient of the considered CDM3Y-Paris-IVF1 EOSs. The corresponding fractions for the maximum stellar mass (${M}_{\text{max}}$) predicted by each EOS are also shown in Fig. \ref{fig:Fig2a}. As revealed in Fig. \ref{fig:Fig2a}, the non-neutron particle fractions exhibit a monotonic decrease with increasing the stiffness of the core ANM, which correlates with increasing the indicated ${M}_{\text{max}}$(${K}_{0}$). For instance, the central ${x}_{pc}$ decreases from 0.376 for ${M}_{\text{max}}$(${K}_{0}$ = 160 MeV) = 0.91 ${\text{M}}_{\odot}$ to 0.077 for ${M}_{\text{max}}$(330 MeV) = 2.4 ${\text{M}}_{\odot}$. The proton fraction exhibits a clear mass dependence at fixed stiffness. As an example, the proton fraction grows from 0.067 (0.7 ${\text{M}}_{\odot}$) to 0.161 at the corresponding maximum mass configuration (2.13 ${\text{M}}_{\odot}$). With a similar decreasing behavior of the electron and muon fractions with ${K}_{0}$, the relative electron-to-proton fraction experiences a mild suppression with growing ANM stiffness, while the corre-sponding muon-to-proton and muon-to-electron ratios display analogous slight decreases. Maintaining fixed stiffness of core ANM, while ${x}_{ec}$, ${x}_{\mu c}$, ${x}_{\mu c}/{x}_{pc}$ and ${x}_{\mu c}/{x}_{ec}$ slightly increase with increasing the NS mass, ${x}_{ec}/{x}_{pc}$ shows an opposite decreasing behavior. \par
\begin{figure*}[!htbp]
	\subfigure[ \label{fig:Fig2c}]{
		\includegraphics[height=13cm,width=1.0\linewidth]{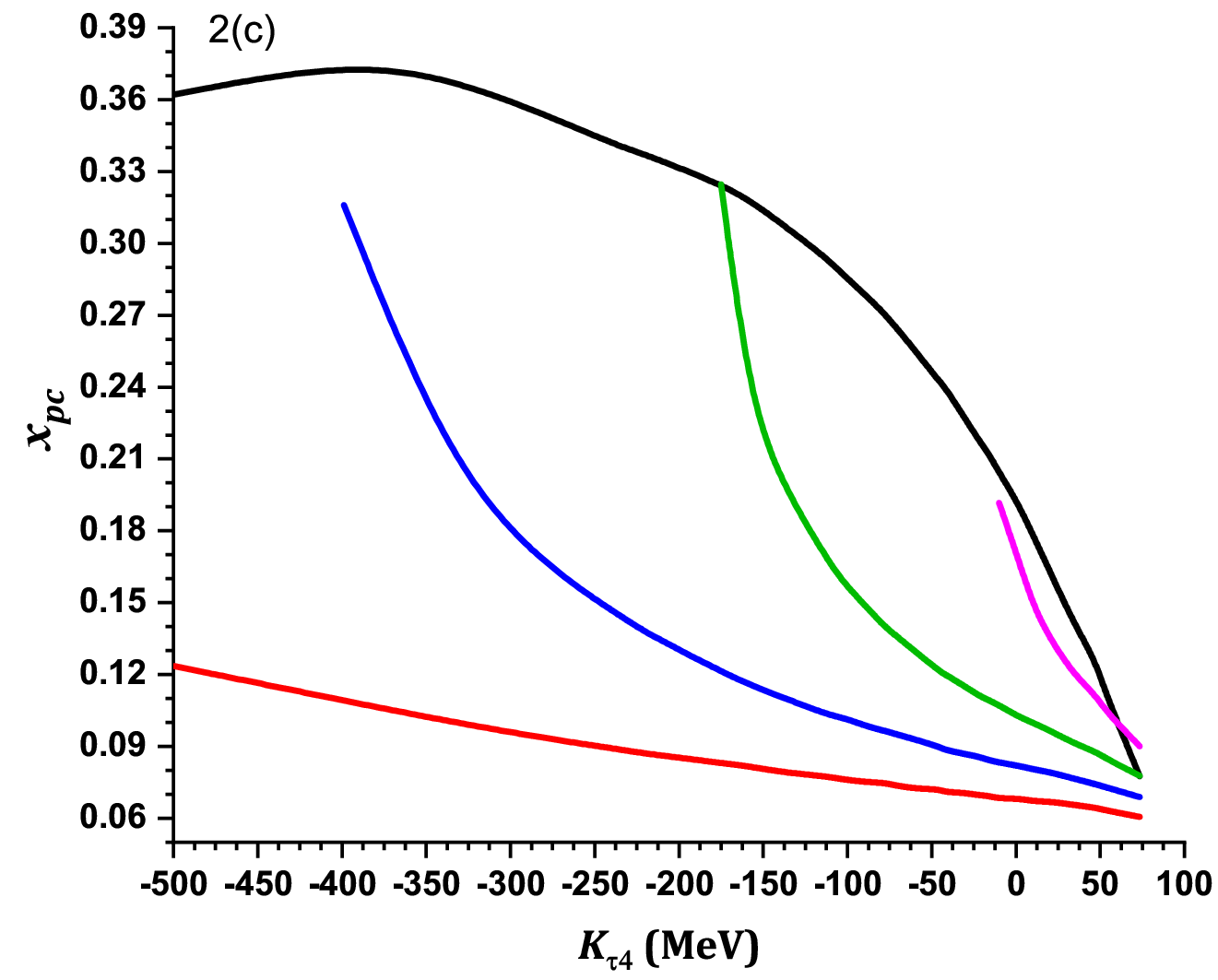}}
	\caption{\label{fig:Figure2} The incompressibility ${K}_{0}$-dependence of the proton (${x}_{pc}$), electron (${x}_{ec}$), and muon (${x}_{\mu c}$) fractions in cold $\beta$-stable $npe\mu$ NS matter, at its core center, for stellar masses 0.7 $\text{M}_{\odot}$, $\text{M}_{\odot}$, 1.4 $\text{M}_{\odot}$, and 2 $\text{M}_{\odot}$, and for the estimated maximum NS mass ${M}_{\text{max}}$(${K}_{0}$), based on the CDM3Y-Paris-IVF1 EOSs, and the central proton fraction as a function of the isobaric curvature coefficients (b) ${K}_{\tau 2}$ and (c) ${K}_{\tau 4}$, for the same stellar masses in panel (a).}  
\end{figure*}

To explore how non-neutron fractions depend on higher symmetry-energy coefficients, we plot in Figs. \ref{fig:Fig2b} and \ref{fig:Fig2c} the central proton fraction versus ${K}_{\tau2}$, which counteracts the ANM stiffness, and versus ${K}_{\tau4}$, which correlates with the stiffness. Figure \ref{fig:Fig2b} reveals that ${x}_{ec}$ increases as the ${K}_{\tau2}$'s negative strength decreases (increasing its positivity), opposite to its behavior with ${K}_{0}$. This consequently predict analogous growth patterns for the non-neutron particle fractions with the ${I}_{0}$, ${G}_{0}$, ${Q}_{2}$, ${K}_{6}$, ${H}_{2}$ and ${K}_{\tau6}$ symmetry energy coefficients, and with decreasing the negative strength of ${K}_{2}$. As shown in Fig. \ref{fig:Fig2c}, ${x}_{ec}$ decreases with decreasing negativity of ${K}_{\tau4}$ towards its positive values mirroring the indicated trend with ${K}_{0}$. A stiffer EOS characterized by more negative ${K}_{\tau2}$ combined with positive ${K}_{\tau4}$ produces a faster fall-off in isovector incompressibility \cite{HashemHassanien2024}. This leads to a more pronounced reduction in the ${x}_{pc}$ compared to the case of less negative ${K}_{\tau2}$ with negative ${K}_{\tau4}$. This indicates a similar decline in the $p$, $e$, and $\mu$ fractions as ${Q}_{4}$ increases and as the negativity of ${Q}_{0}$, ${H}_{0}$, ${I}_{2,4}$, ${G}_{2}$, and ${K}_{4}$ symmetry-energy coefficients decreases. This rise in the proton-fraction with increasing ${Q}_{2}$ and more negative ${Q}_{0}$ can be also indicated by the calculations in Ref. \cite{MalikPais2024}, and its increase with ${K}_{2}$ has been indicated through a meta-modeling approach applied to NS matter, utilizing relativistic density functional \cite{CharMondal2023}. As the NS mass increases, the $p$, $e$, $\mu$ fractions grow in its core, showing a stronger dependence on the stiffness of the ANM and its related higher-order symmetry-energy parameters. Increasing the non-neutron particle fractions softens the EOS, leading to lower predicted NS masses and enhanced symmetry pressure \cite{SeifHashem2025}. It also alters the NS's chemical composition, which governs key transport properties like thermal and electrical conductivity \cite{CharMondal2023}.
\subsection{\label{sec:level3c} Direct Urca density, pressure, and proton fraction}
\begin{figure*}[!htbp] 
	\centering	
	\subfigure[ \label{fig:Fig3a}]{
		\includegraphics[height=12cm,width=1.0\linewidth]{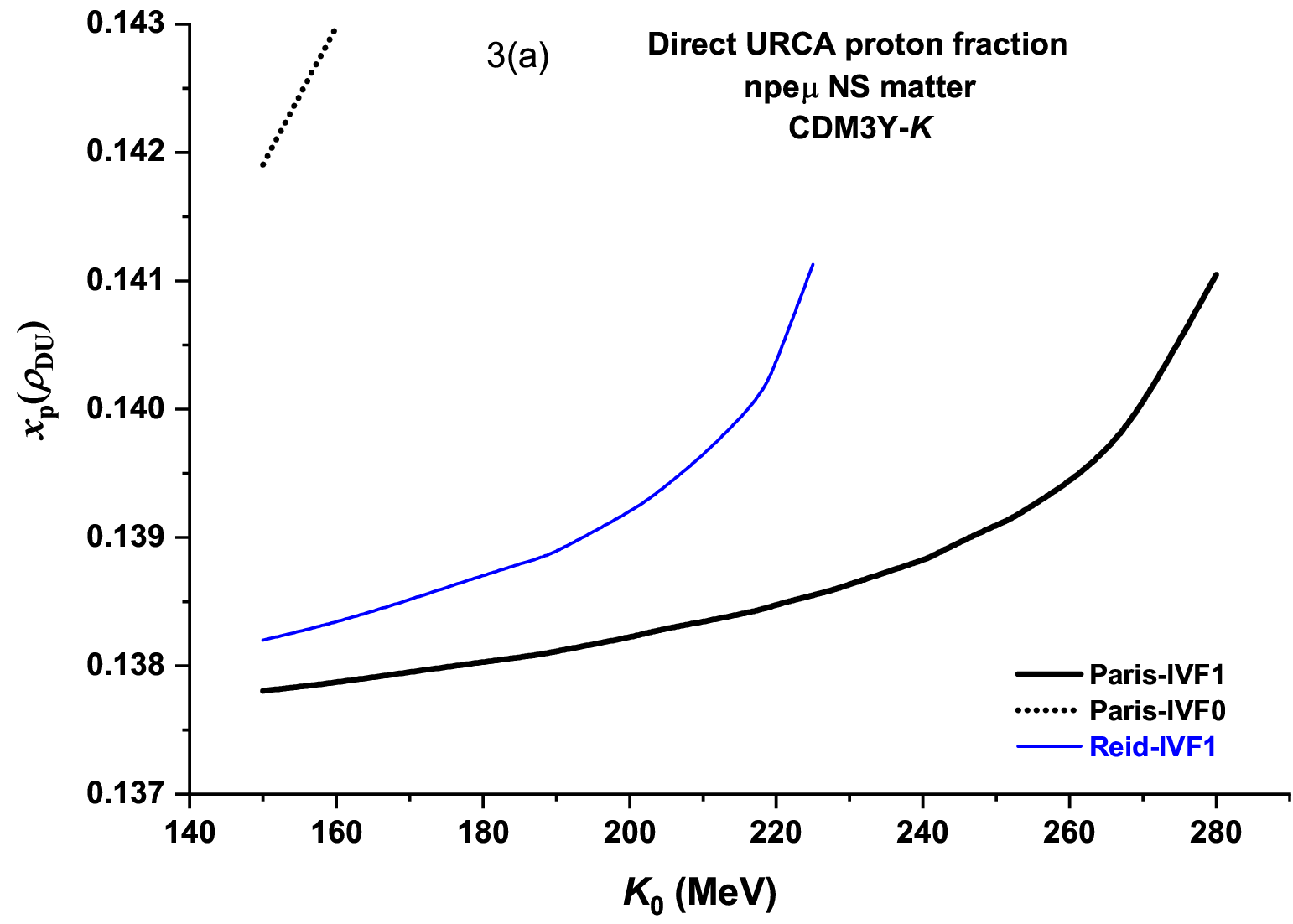}}
	~
	\centering
	\subfigure[ \label{fig:Fig3b}]{
		\includegraphics[height=12cm,width=1.0\linewidth]{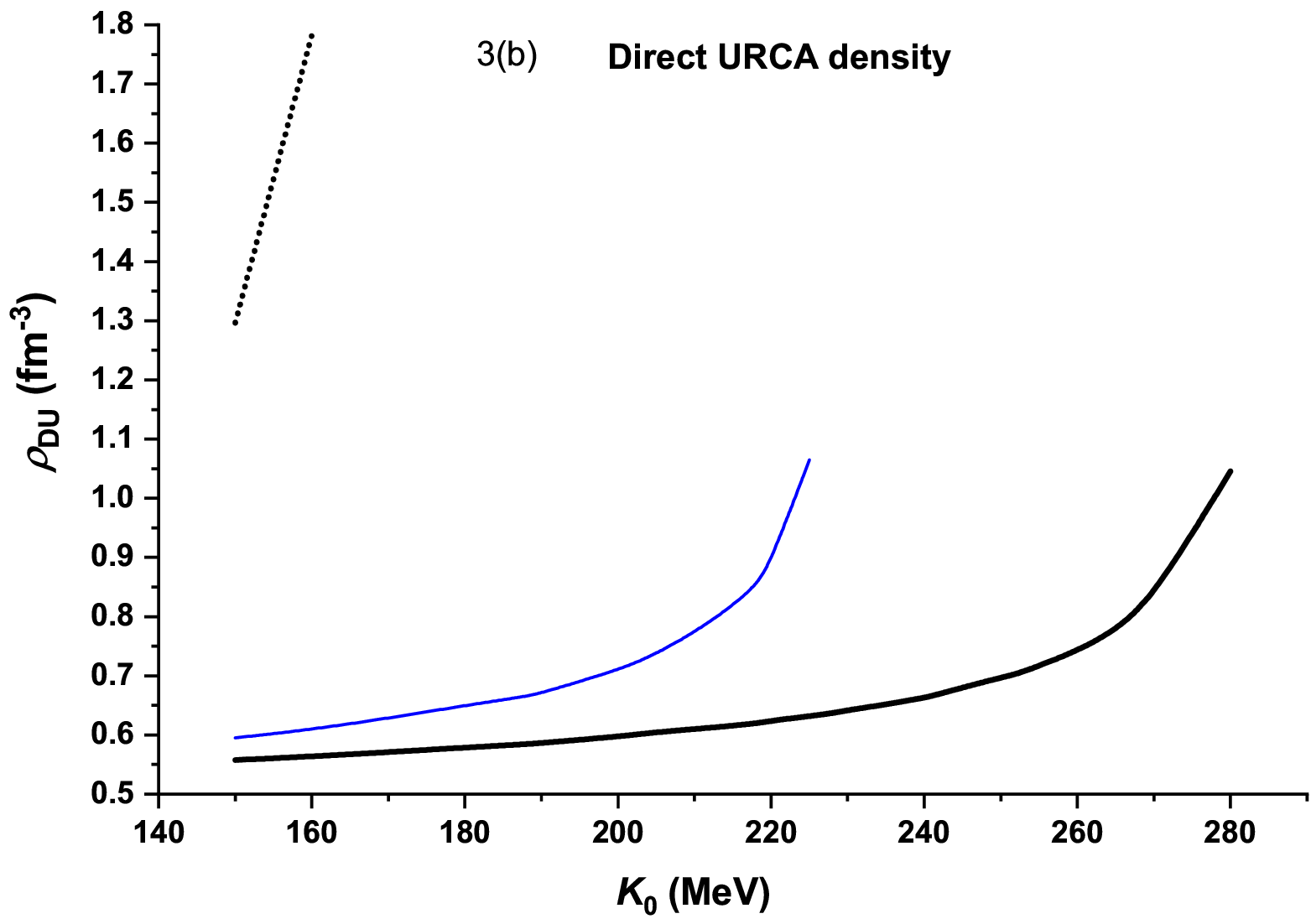}}	
\end{figure*}

\begin{figure*}[!htbp] 
	\centering	
	\subfigure[ \label{fig:Fig3c}]{
		\includegraphics[height=12cm,width=1.0\linewidth]{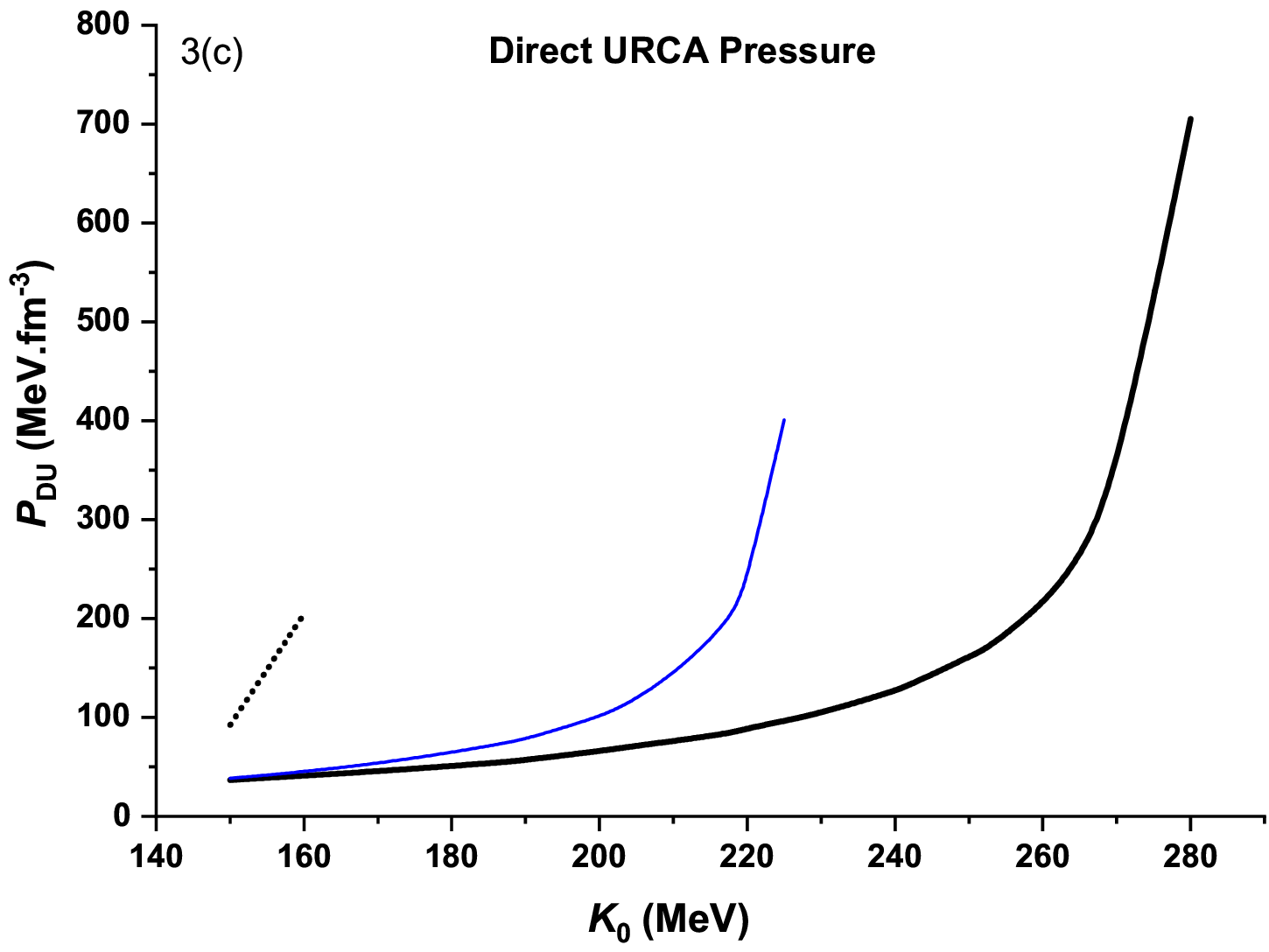}}
	~
	\centering
	\subfigure[ \label{fig:Fig3d}]{
		\includegraphics[height=12cm,width=1.0\linewidth]{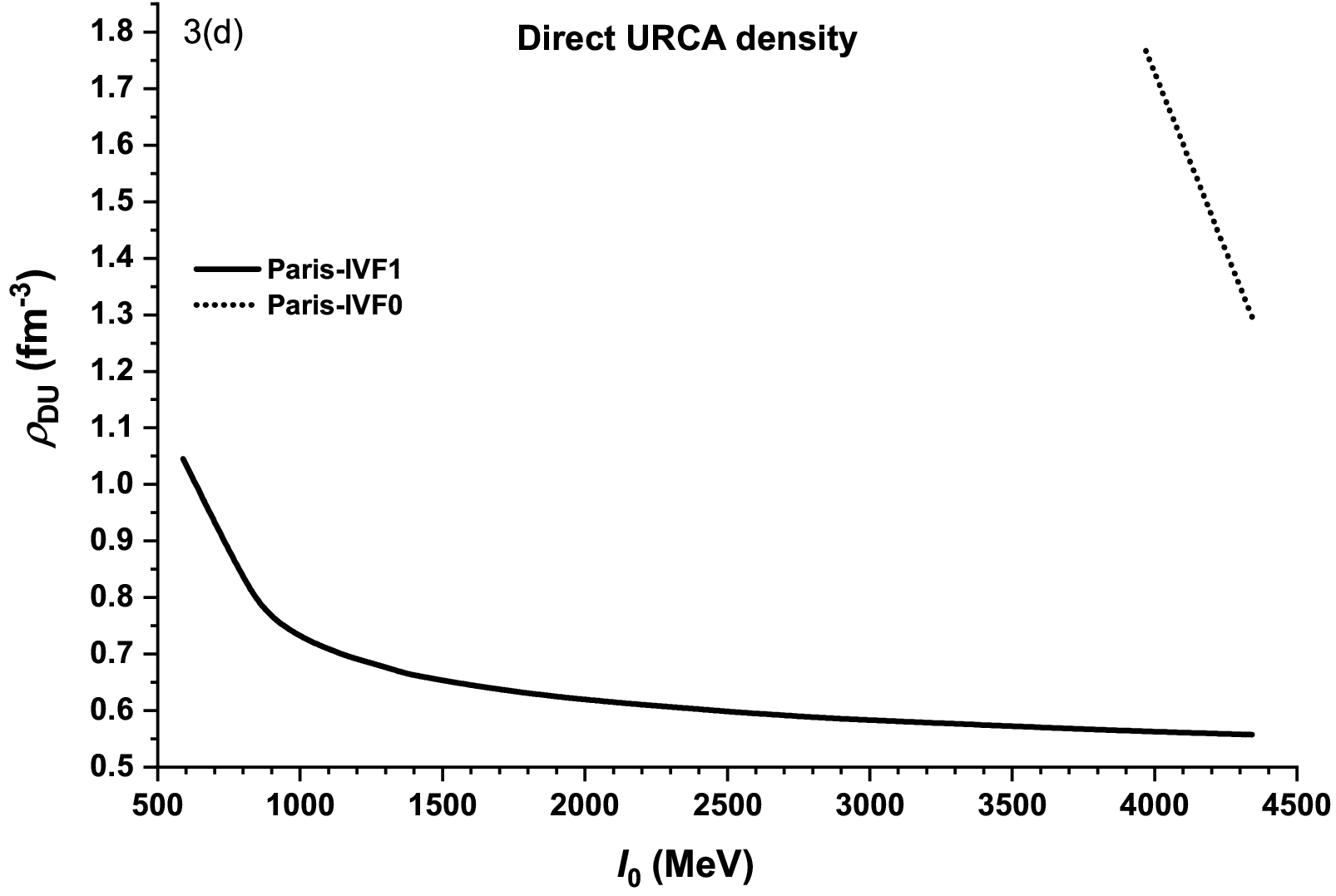}}	
\end{figure*}

The proton fraction is a crucial parameter that governs the NS neutrino emission and its thermal evolution, dictating whether its cooling proceeds rapidly through direct Urca (DU) neutrino-cooling mechanism, which involves nucleon decay and electron capture, or slowly through modified Urca or cooper pair breaking/formation mechanisms \cite{PageLattimer2004,LimHolt2021}. The EOS of core NS matter and its symmetry energy significantly influence such competition \cite{HashemHassanien2024,DasWei2024}. The threshold proton-fraction required for the direct Urca process given by Eq. (\ref{eqn:xDU}), ${x}_{\text{DU}}={x}_{p}$(DU),  is plotted in Fig. \ref{fig:Fig3a} as a function of the isoscalar incompressibility ${K}_{0}$ of NS core $npe\mu$ matter, calculated using the CDM3Y-$K$  EOSs derived from the Paris-IVF1, Paris IVF0, and Reid–IVF1 effective NN interactions. The CDM3Y-Reid-IVF0 EOSs do not predict a sufficient ${x}_{\text{DU}}$ to allow DU process to take place \cite{HashemHassanien2024}. The corresponding DU density, ${\rho}_{\text{DU}}=\rho({x}_{\text{DU}})$, and pressure, ${P}_{\text{DU}}=P({\rho}_{\text{DU}},{x}_{\text{DU}})$ are displayed in Figs. \ref{fig:Fig3b} and \ref{fig:Fig3c}, respectively. \par
\begin{figure*}[!htbp]
	\subfigure[ \label{fig:Fig3e}]{
		\includegraphics[height=13cm,width=1.0\linewidth]{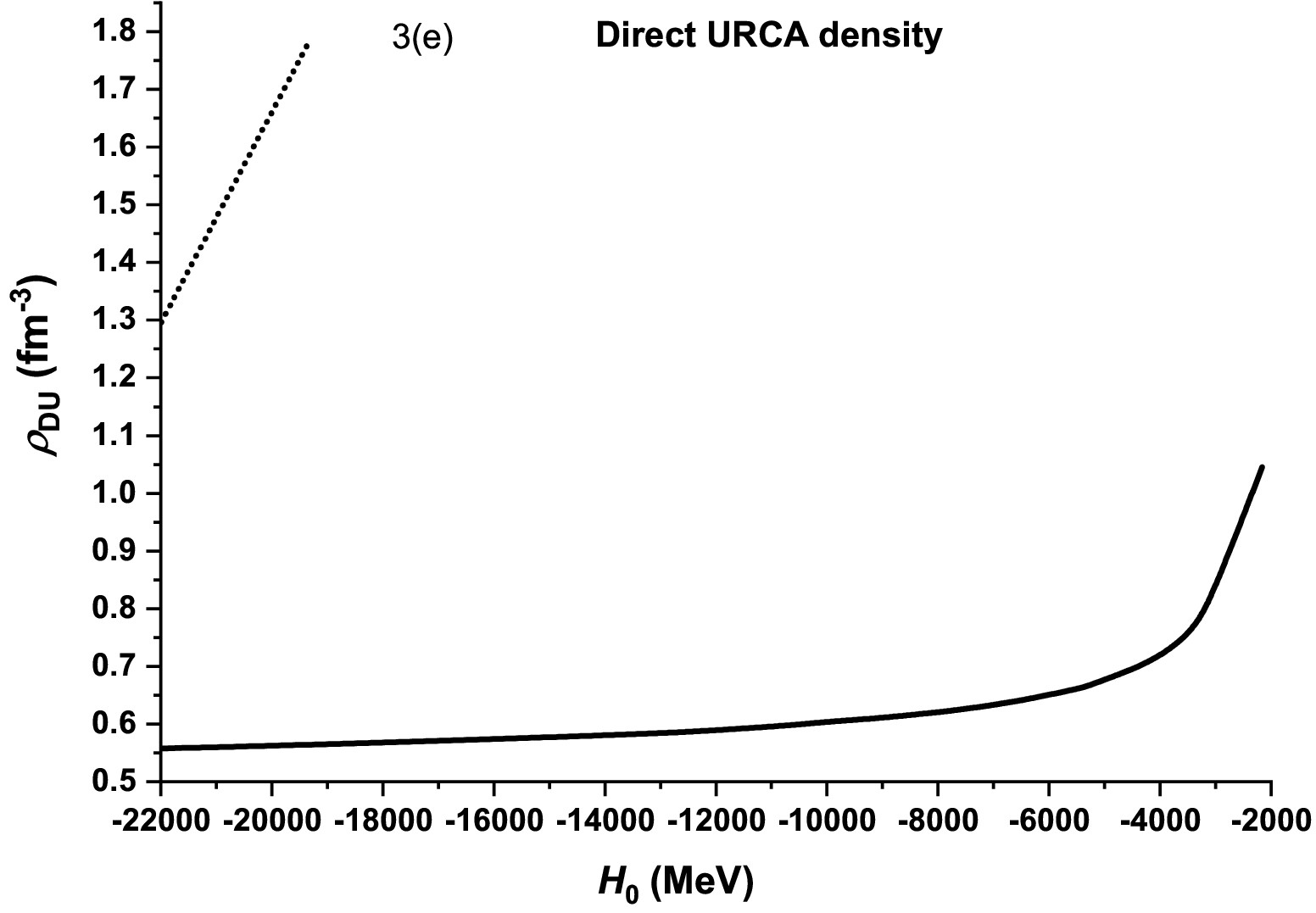}}
	\caption{\label{fig:Figure3} Threshold (a) proton fraction, (b) density, and (c) pressure required to initiate the DU process (Eq. (\ref{eqn:xDU})) as a function of ${K}_{0}$, based on the CDM3Y-Paris-IVF1 (IVF0) and CDM3Y-Reid-IVF1 EOSs, and the threshold density ${\rho}_{\text{DU}}$ as a function of the isoscalar (d) kurtosis ${I}_{0}$ and (e) fifth-order ${H}_{0}$ symmetry-energy coefficients, based on the CDM3Y-Paris-IVF1 (IVF0) EOSs.}  
\end{figure*}

Figure \ref{fig:Figure3} demonstrates that the isovector density dependence IVF1, which predicts stiff symmetry energy at high densities \cite{SeifHashem2023}, yields a lower threshold DU proton fraction, density, and pressure, compared to the IVF0 density dependent associated with softer symmetry energy at supra-saturation densities and less density-slope $L$ \cite{HashemHassanien2024}. The IVF1 density dependence also predicts a broader range of stiff NM that supports the DU process, whereas the soft high-density symmetry energy of IVF0 indicates only extremely soft NM for DU. Similarly, CDM3Y-Paris-IVF1 EOSs, with stiffer symmetry energy than CDM3Y-Reid-IVF1, predict less ${x}_{\text{DU}}$, ${\rho}_{\text{DU}}$ and ${P}_{\text{DU}}$. Over the ascending range ${K}_{0}=150-280$ MeV, the CDM3Y-Paris-IVF1 EOSs predict an increasing DU threshold density ${\rho}_{\text{DU}}= (3.28-6.15){\rho}_{0}$, accompanied by a rise in both the proton fraction (${x}_{\text{DU}}= 0.138-0.141$) and pressure (${P}_{\text{DU}}=37-705$ MeV ${\text{fm}}^{-3}$). This correlates with the reduction in the $\beta$-stable proton-fraction at higher densities in stiffer core NM \cite{HashemHassanien2024}, which increases the threshold ${\rho}_{\text{DU}}$ and its corresponding ${P}_{\text{DU}}$.  The stiffer CDM3Y-Paris-IVF1 EOSs (with ${K}_{0} > 180$ MeV) do not support DU process, where ${x}_{p}$ falls below the necessary threshold. Comparing the stiffness dependence of ${x}_{\text{DU}}$ in Fig. \ref{fig:Fig3a} with that of ${x}_{pc}$ in Fig. \ref{fig:Fig2a}, we observe that this dependence strengthens for ${x}_{pc}$ at higher densities. In both figures, the variation in these quantities also grows at larger values. Over the considered stiffness range, ${x}_{pc}$ spans ($0.46-2.21$)${\rho}_{0}$, while ${x}_{\text{DU}}$ varies only slightly within ($0.81-0.83$)${\rho}_{0}$. This suggests an increasing role of higher-order symmetry energy coefficients at ultrahigh densities, where their influences can be finely tuned. \par
Based on the above-mentioned correlations between symmetry-energy coefficients and NM stiffness, increasing ${Q}_{4}$ and reducing the negativity of ${Q}_{0}$, ${H}_{0}$, ${I}_{2,4}$, ${G}_{2}$, ${K}_{4}$, and ${K}_{\tau4}$ will enhance ${x}_{\text{DU}}$ and ${\rho}_{\text{DU}}$, as well as their corresponding ${P}_{\text{DU}}$, whereas these quantities decrease with larger ${I}_{0}$, ${G}_{0}$, ${Q}_{2}$, ${K}_{6}$, ${H}_{2}$ and ${K}_{\tau6}$ coefficients, or with diminished negativity of ${K}_{2}$ and ${K}_{\tau2}$. Figures \ref{fig:Fig3d} and \ref{fig:Fig3e} confirm this by showing how the minimum ${\rho}_{\text{DU}}$, allowing the DU process, varies with the isoscalar kurtosis (${I}_{0}$) and ${H}_{0}$ coefficients, respectively. As expected, ${\rho}_{\text{DU}}$ decreases with larger ${I}_{0}$ and increases as ${H}_{0}$ becomes less negative. The reduction of ${\rho}_{\text{DU}}$ with weakening ${K}_{2}$ negativity has been indicated also across different relativistic mean-field EOSs \cite{BoukariMalik2024}. A lower threshold ${\rho}_{\text{DU}}$ along with reduced ${x}_{\text{DU}}$ accelerate cooling \cite{LimSchwenk2024}, generating observable neutrino signals from lighter NS, while increasing them restricts the direct Urca cooling mechanism to only the massive NS.
\subsection{\label{sec:level3d} Moment of inertia of NS and its crustal fraction}
\begin{figure*}[!htbp] 
	\centering	
	\subfigure[ \label{fig:Fig4a}]{
		\includegraphics[height=12cm,width=1.0\linewidth]{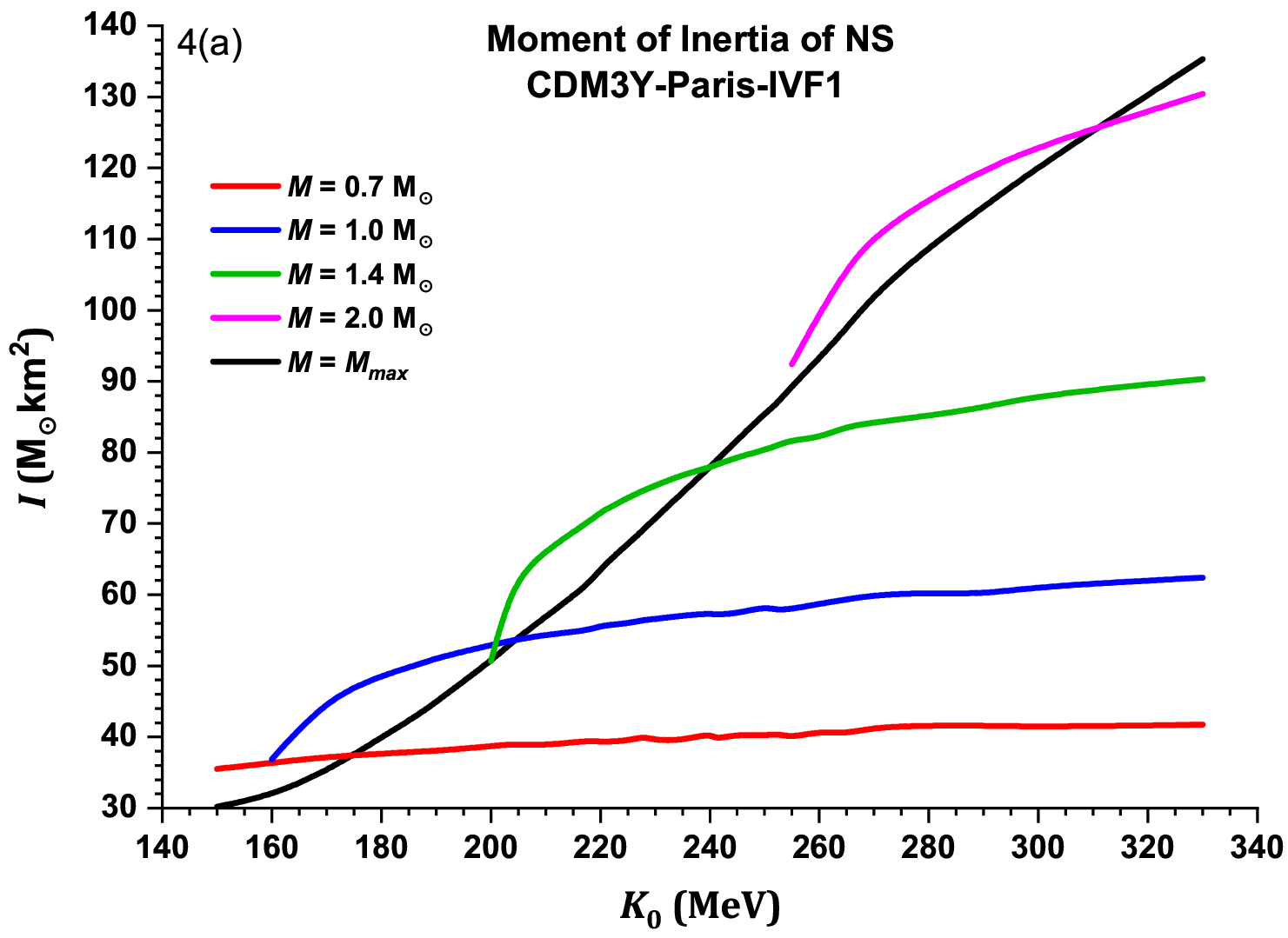}}
	~
	\centering
	\subfigure[ \label{fig:Fig4b}]{
		\includegraphics[height=12cm,width=1.0\linewidth]{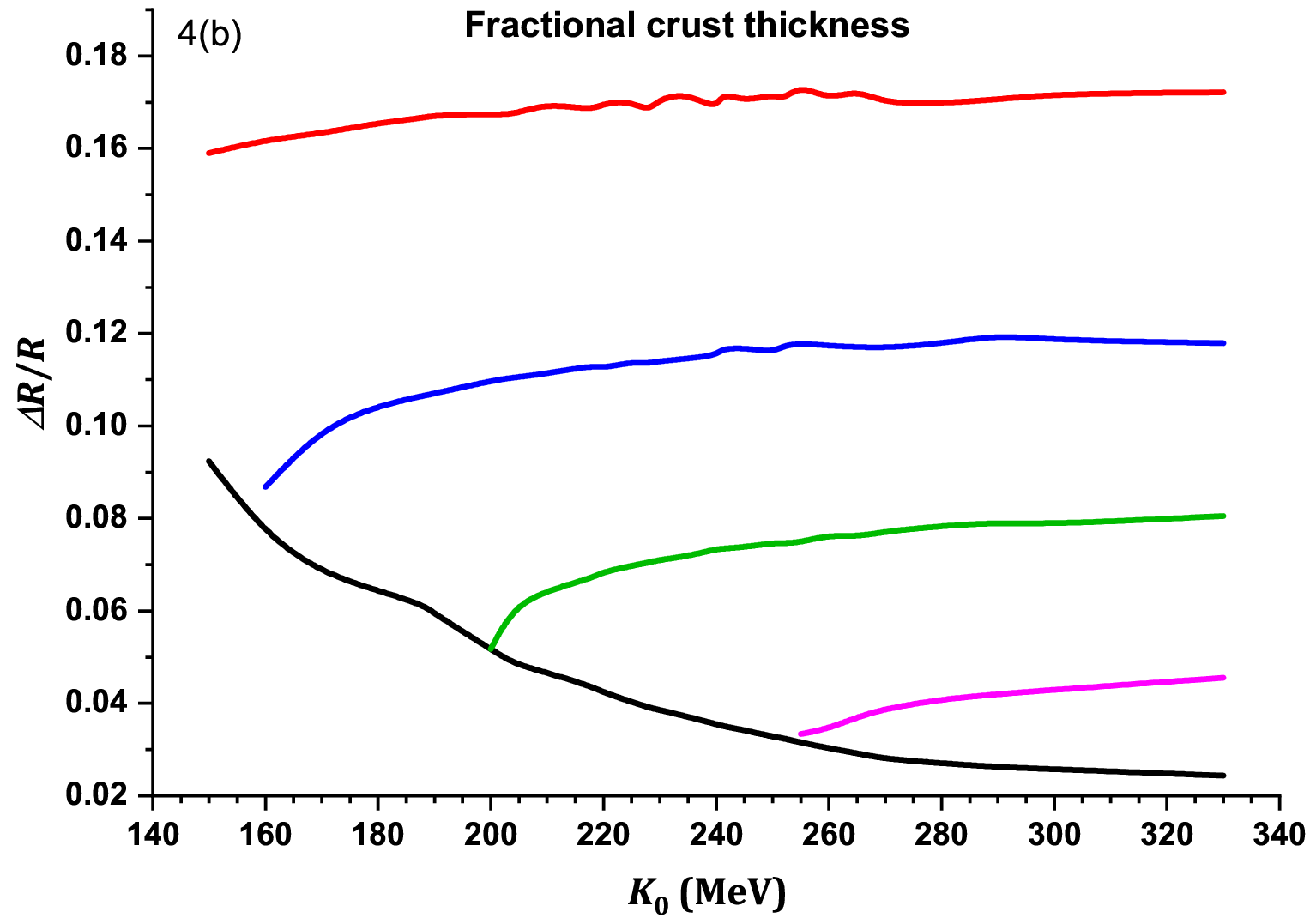}}	
\end{figure*}

We proceed to examine how the stiffness of the NM and the symmetry-energy coefficients influence the MI of NS’s for varying masses, including the crust’s contribution to the moment of inertia. This is under the consideration of distinct EOSs for the bound clusters and heavy nuclei present in the inner and outer crustal regions. In Fig. \ref{fig:Fig4a}, we present the computed MI for slowly rotating NSs of masses 0.7 $\text{M}_{\odot}$, $\text{M}_{\odot}$, 1.4 $\text{M}_{\odot}$ and 2 $\text{M}_{\odot}$, along with the maximum mass (${M}_{\text{max}}$) supported by the employed EOS, plotted against the SNM incompressibility characterizing the considered EOS. The EOSs predicting NS masses of $\text{M}_{\odot}$, 1.4 $\text{M}_{\odot}$ and 2 $\text{M}_{\odot}$ show larger ${K}_{0}$ than 160 MeV, 200 MeV and 255 MeV, respectively, whereas the softer EOSs cannot support NS mass larger than $\text{M}_{\odot}$. The influence of NM stiffness in the core regions on the MI of NS's grows progressively stronger with increasing gravitational mass, as demonstrated in Fig. \ref{fig:Fig4a}. The changing in the behavior of the MI for NS of a given mass, relative to that of the indicated ${M}_{\text{max}}$, according to the stiffness of the employed EOS emerge from competing mass, radius and compactness variations with the EOS. For instance, both $R$(2 $\text{M}_{\odot}$) and $R$(${M}_{\text{max}}$) increase with increasing the incompressibility of the employed EOS from ${K}_{0}$ = 270 MeV to 330 MeV, by about 1.04 km and 0.77 km, respectively, mean-while the corresponding compactness of the NS decreases for the NS (2 $\text{M}_{\odot}$) by about $-0.021$ but it increases for the NS (${M}_{\text{max}}$) by about 0.016. This leads to an increase in the relative $I(2\,{\text{M}}_{\odot})/I({M}_{\text{max}})$ for the softer EOS (270 MeV), while it diminishes under the stiffer EOS (330 MeV). \par
\begin{figure*}[!htbp] 
	\centering	
	\subfigure[ \label{fig:Fig4c}]{
		\includegraphics[height=12cm,width=1.0\linewidth]{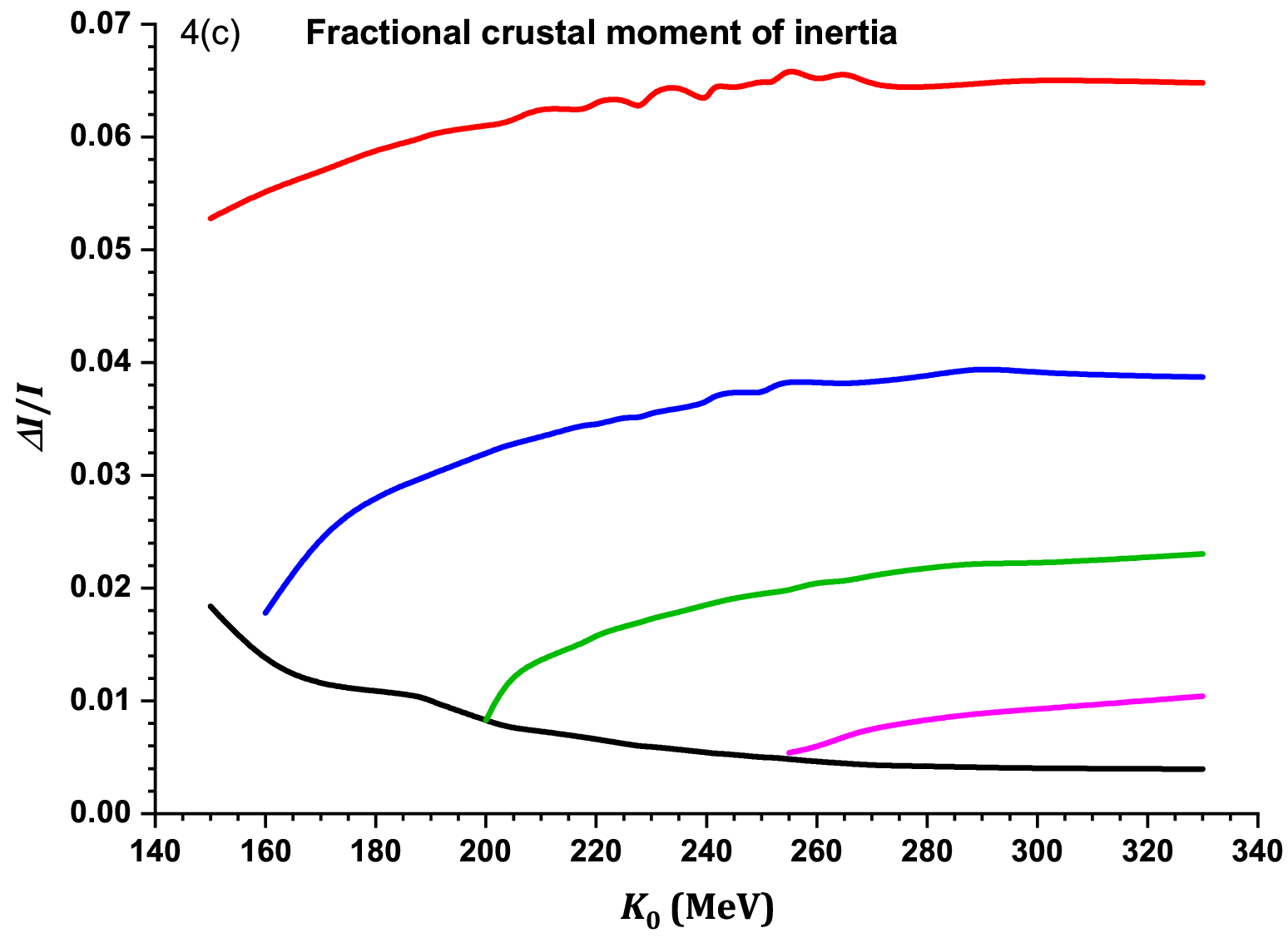}}
	~
	\centering
	\subfigure[ \label{fig:Fig4d}]{
		\includegraphics[height=12cm,width=1.0\linewidth]{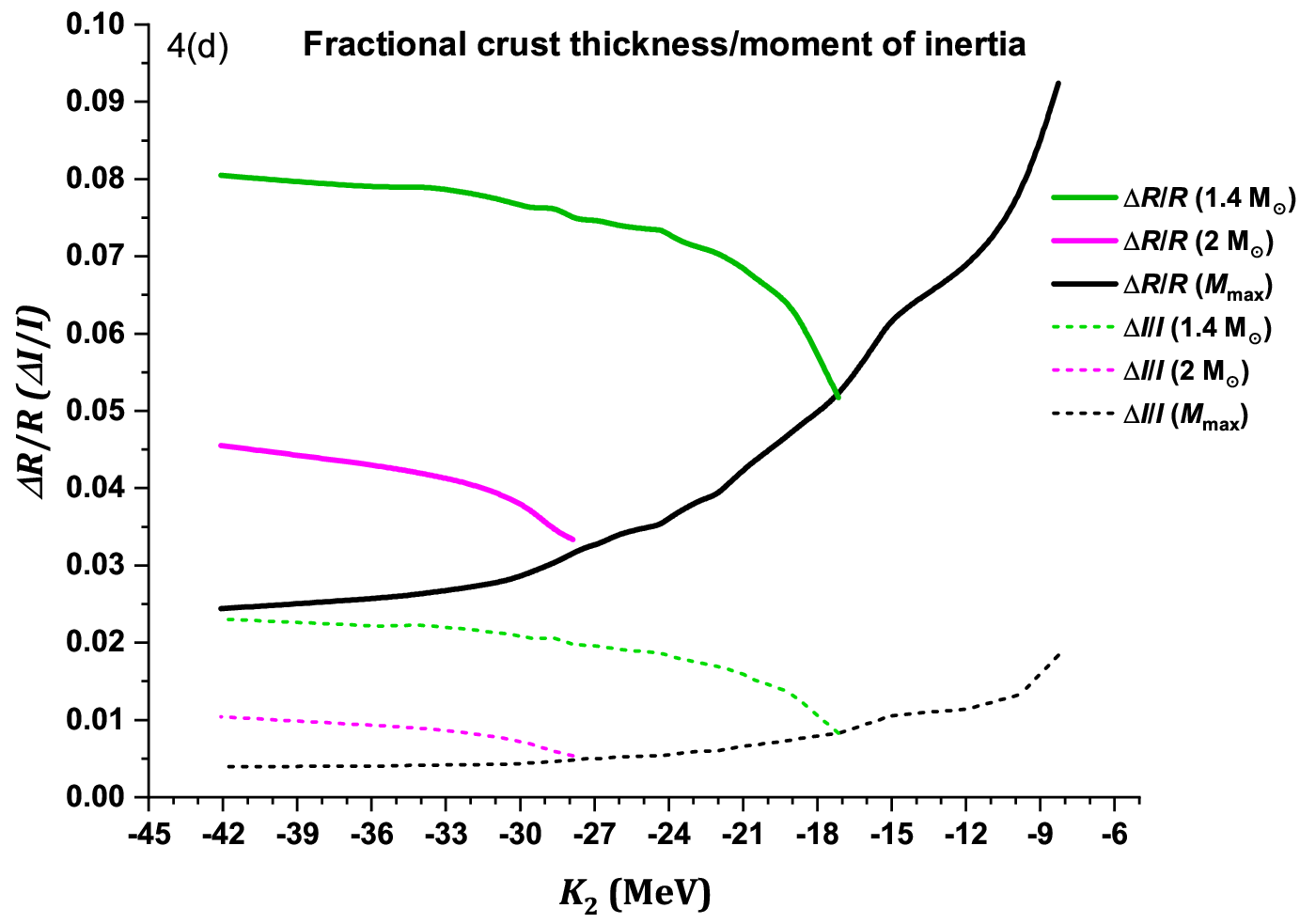}}	
\end{figure*}

The primary parameter characterizing the NS crust is its thickness relative to the stellar radius, shown in Figure \ref{fig:Fig4b} as a function of NM incompressibility. The fractional crust thickness demonstrates a clear anti-correlation with neutron star mass, but minimal sensitivity to core matter stiffness, as shown in Fig. \ref{fig:Fig4b}. This stiffness dependence becomes even weaker for increasingly hard core EOS. The minimum fractional crust thickness of ${M}_{\text{max}}$(${K}_{0}$) exhibits a stronger, opposite trend, showing significant decrease with increasing incompressibility. Given that $\Delta I/I \ge 0.014$ is necessary to explain glitches in the Vela pulsar \cite{LinkEpstein1999}, our results show that this criterion is fulfilled for $M \le 1.6\, {\text{M}}_{\odot}$. Accounting for mass variations, the crustal fraction of moment of inertia, demonstrated in Fig. \ref{fig:Fig4c}, largely mirrors the fractional crust thickness behavior with ${K}_{0}$, but exhibits stronger core stiffness dependence at fixed NS mass and weaker dependence for that of ${M}_{\text{max}}$(${K}_{0}$). \par
\begin{figure*}[!htbp]
	\subfigure[ \label{fig:Fig4e}]{
		\includegraphics[height=13cm,width=1.0\linewidth]{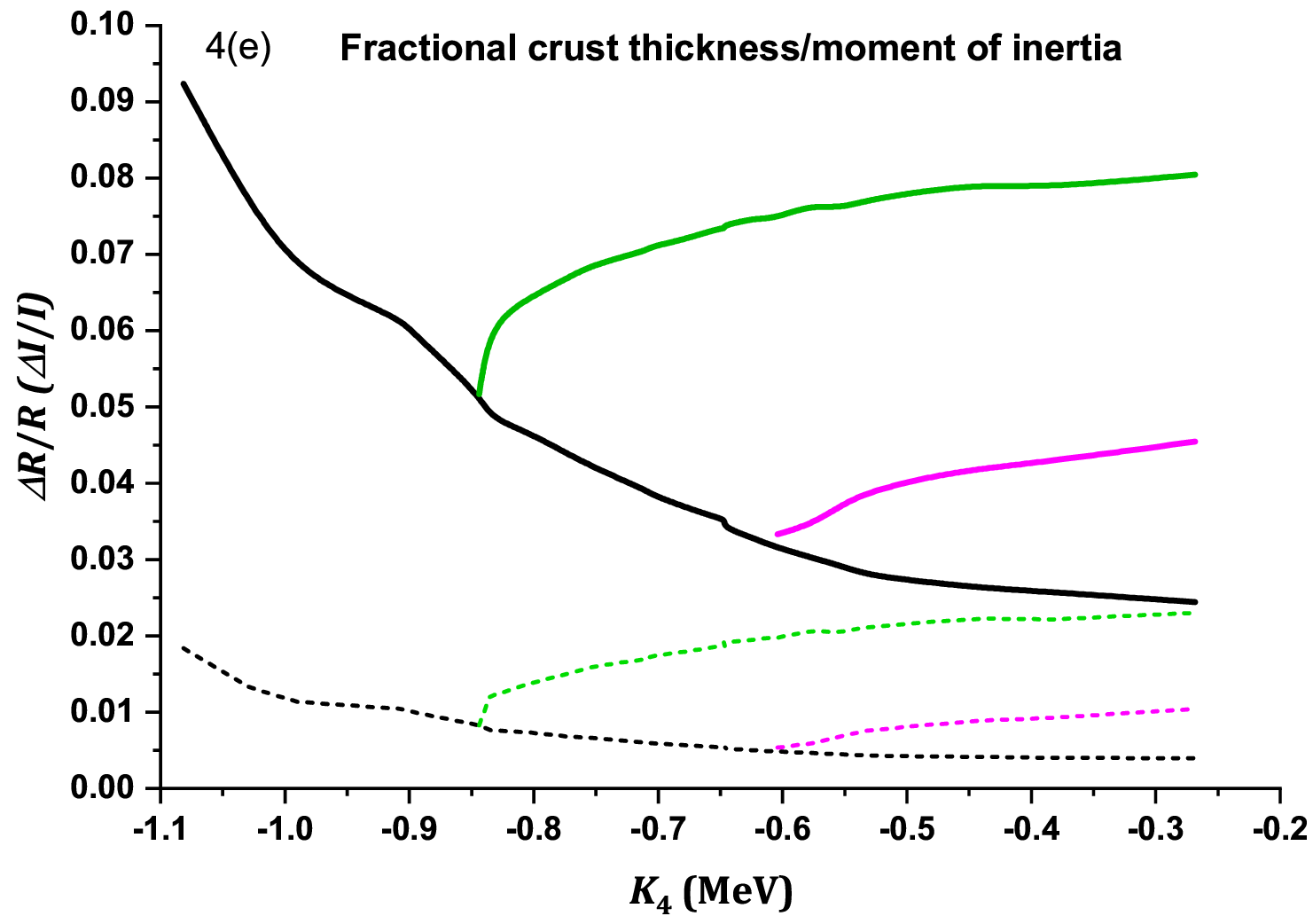}}
	\caption{\label{fig:Figure4} (a) Total moment of inertia, (b) fractional crustal thickness relative to stellar radius, and (c) crustal fraction of the moment of inertia of slowly rotating NS with masses 0.7 $\text{M}_{\odot}$, $\text{M}_{\odot}$, 1.4 $\text{M}_{\odot}$ and 2 $\text{M}_{\odot}$, that having the estimated ${M}_{\text{max}}$(${K}_{0}$), as functions of ${K}_{0}$, based on the CDM3Y-Paris-IVF1 EOSs, and the fractional crustal thickness as a function of the isoscalar curvature coefficient (d) ${K}_{2}$ and (e) ${K}_{4}$.}  
\end{figure*}

In view of the above indicated correlations between the different symmetry coefficients and the stuffiness of NM, an increase in the MI of NS and in the fractional crust thickness (moment of inertia), but a decrease in the fractional crust thickness and crustal moment of inertia fraction for the estimated ${M}_{\text{max}}$(${K}_{0}$) configuration, are expectedly correlate with larger ${Q}_{4}$, less negative ${Q}_{0}/{H}_{0}/{I}_{2,4}/{G}_{2}/{K}_{4,\tau 4}$, smaller ${I}_{0}/{G}_{0}/{Q}_{2}/{K}_{6,\tau 6}/{H}_{2}$, and more negative ${K}_{2,\tau 2}$. To show examples for this, we plot in Fig. \ref{fig:Fig4d} (Fig. \ref{fig:Fig4e}) $\Delta R/R$ and $\Delta I/I$ for NS of masses 1.4 ${\text{M}}_{\odot}$, 2 ${\text{M}}_{\odot}$, and ${M}_{\text{max}}$(${K}_{0}$), as functions of incompressibility ${K}_{2}$ (${K}_{4}$) coefficient, which opposes (follows) the stiffening behavior of ANM. Figure \ref{fig:Fig4d} (Figure \ref{fig:Fig4e}) confirms the decreasing (increasing) behavior of both $\Delta R/R$ and $\Delta I/I$ for NS of a given mass with decreasing the negativity of ${K}_{2}$ (${K}_{4}$), while they increase (decrease) for ${M}_{\text{max}}$(${K}_{0}$) with it, showing opposite (simulating) trend to their dependence on ${K}_{0}$. The currently results align with those obtained for the behavior with ${K}_{2}$ derived based on an ensemble of unified NS crust and core EOSs, constructed using an extended Skyrme energy density functional and two piecewise polytropes at higher densities, with constraints from PREX and NICER+LIGO/VIRGO/Electromagnetic (NICER+LV/EM) astrophysical data \cite{NewtonBalliet2022}. The best agreement can be seen for the NICER+LV/EM indicated values of $\Delta I/I={0.037}_{-0.015}^{+0.018}$(${K}_{2}={-39}_{-111}^{+148}$ MeV) and its corresponding $\Delta R/R={0.105}_{-0.023}^{+0.019}$, for NS (1.4 ${\text{M}}_{\odot}$) \cite{NewtonBalliet2022}. The decreasing trend of $I$, $\Delta I/I$, and $\Delta R/R$ with ${K}_{2}$ can be also seen from the change in their inferred values with the corresponding median values of ${K}_{2}$ within the range from $-76$ to $-2$ MeV. The Negele-Vautherin EOSs \cite{ZhangLi2025} suggest similar decreasing behavior of $\Delta I/I$ as ${K}_{2}$ becomes less negative and with decreasing ${Q}_{0}$. The changes in the NS crustal thickness and its MI respond directly to the estimated transition density. Increasing (decreasing) them corresponds to earlier (delayed) core-crust transition, at higher (lower) transition density inside the core. This correlation reflects how both the transition density (Fig. \ref{fig:Figure1}) and these structural parameters (Fig. \ref{fig:Figure4}) respond similarly to variations in the EOS stiffness and in the symmetry-energy coefficients.

\subsection{\label{sec:level3e} Moment of inertia and compactness}
\begin{figure*}[!htbp]
	\subfigure[ \label{fig:Fig5a}]{
		\includegraphics[height=12cm,width=1.0\linewidth]{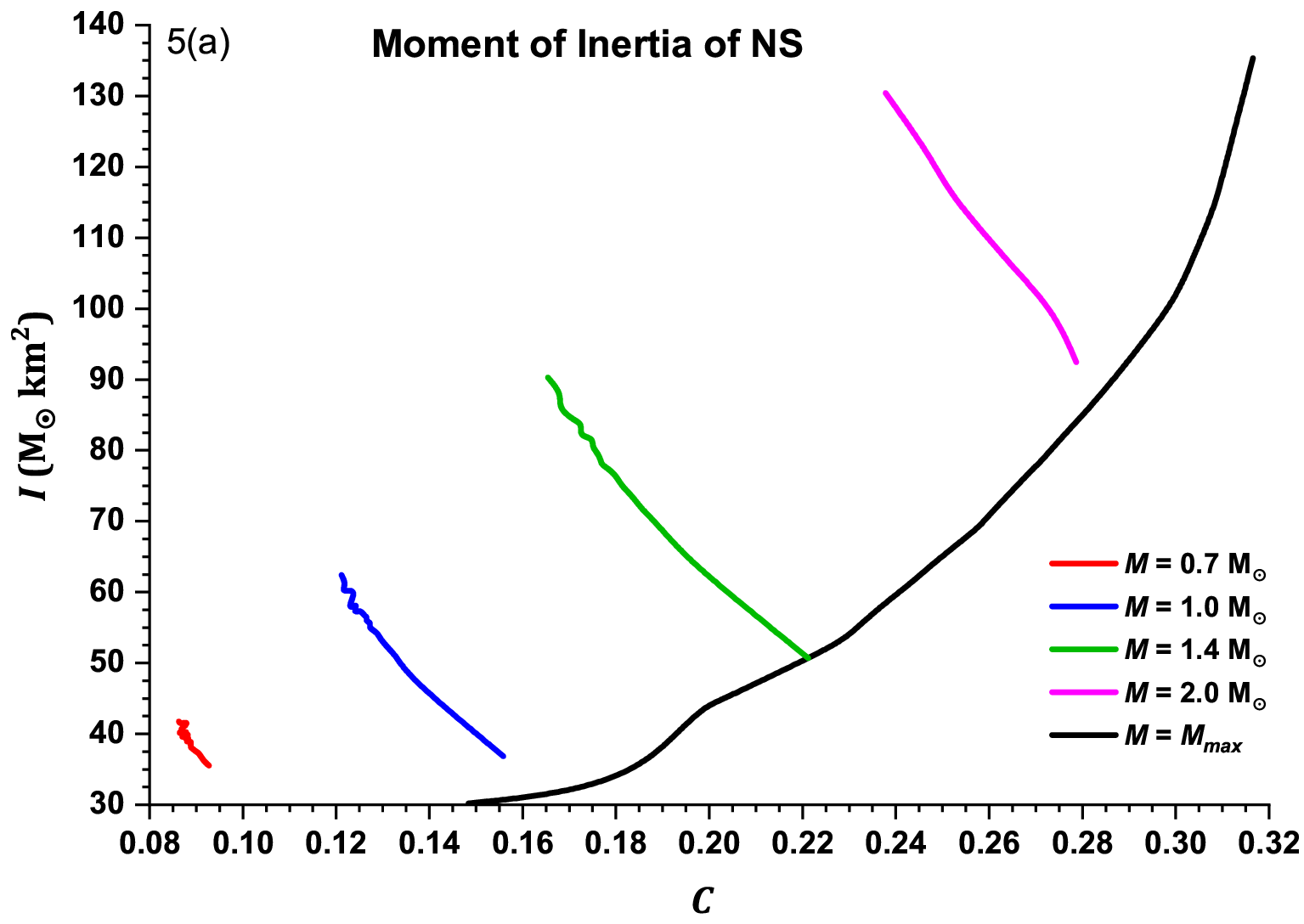}}
	~
	\subfigure[ \label{fig:Fig5b}]{
		\includegraphics[height=12cm,width=1.0\linewidth]{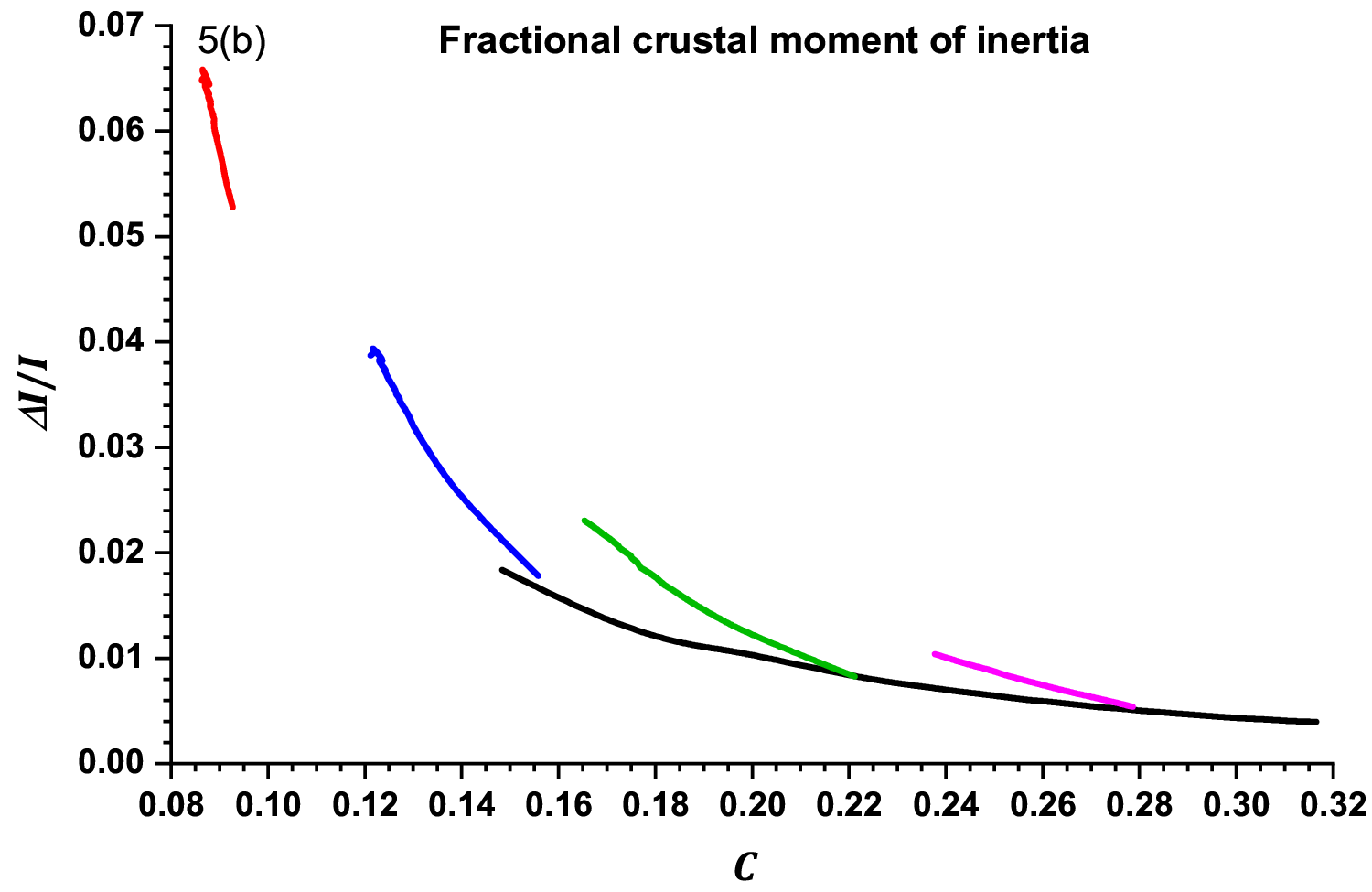}}	
	\caption{\label{fig:Figure5} (a) Total MI and (b) crustal fraction of the moment of inertia, for the same NS masses and based on the same EOSs in Fig. \ref{fig:Figure4}, as a function of the NS compactness.}
\end{figure*}

A neutron star's MI is determined by its compactness, as both are directly related to the NS mass-radius relation. The dependence of the MI (NS) and its crustal fraction on compactness ($\mathpzc{C}$) is illustrated in Figs. \ref{fig:Fig5a} and \ref{fig:Fig5b}, respectively, for the same NS masses considered in Fig. \ref{fig:Fig4a}. Figure \ref{fig:Figure5} reveals that both $I$ and $\Delta I/I$ decrease linearly with $\mathpzc{C}$ in a similar fashion, regardless increasing $\mathpzc{C}$ with the NS mass that produces distinct ranges of $\mathpzc{C}(M)$, with the decreasing linear trend of $\Delta I/I$ with $\mathpzc{C}$ flattens with increasing NS mass and its range of compactness. This correlation reflects both the increasing stiffness of the corresponding EOSs and the associated increase (decrease) in the positivity of their correlated (anti-correlated) symmetry energy coefficients, leading to an overall increasing (decreasing) trend of $I$ ($\Delta I/I$) with growing compactness of the corresponding ${M}_{\text{max}}$(${K}_{0}$). Multiple studies \cite{PopchevStaykov20219,StaykovDoneva2016} have demonstrated the universality of the MI and compactness relation, within General Relativity and other frameworks. Such universality is expected to arise from compensating effects in the EOS, such as the increases the mass and radius with a simultaneous decrease in the central density ${\rho}_{c}$, see Fig. \ref{fig:Figure3} in \cite{SeifHashemJPG2024}. This maintains the relationship across different EOS models, as demonstrated by the nearly identical mass-radius curve shapes in the region following the indicated maximum radius limit Fig. \ref{fig:Figure1} in \cite{SeifHashemJPG2024}. Meanwhile, for stiffest EOSs, the increased compactness at higher masses compresses the crust to a nearly minimal thickness (Fig. \ref{fig:Fig4b}), resulting in a remarkably consistent $\Delta I/I$ over a wide stiffness range, ${K}_{0}>250$ MeV (Fig. \ref{fig:Fig4c}). Consequently, the EOS contributes minimally to rotational dynamics. The requirement $\Delta I/I\ge$0.014 for Vela pulsar glitches restricts NS compactness in our analysis to be less than 0.21 ($M\le$ 1.6 ${\text{M}}_{\odot}$).

\subsection{\label{sec:level3f} The adiabatic index}
\begin{figure*}[!htbp]
	\subfigure[ \label{fig:Fig6a}]{
		\includegraphics[height=12cm,width=1.0\linewidth]{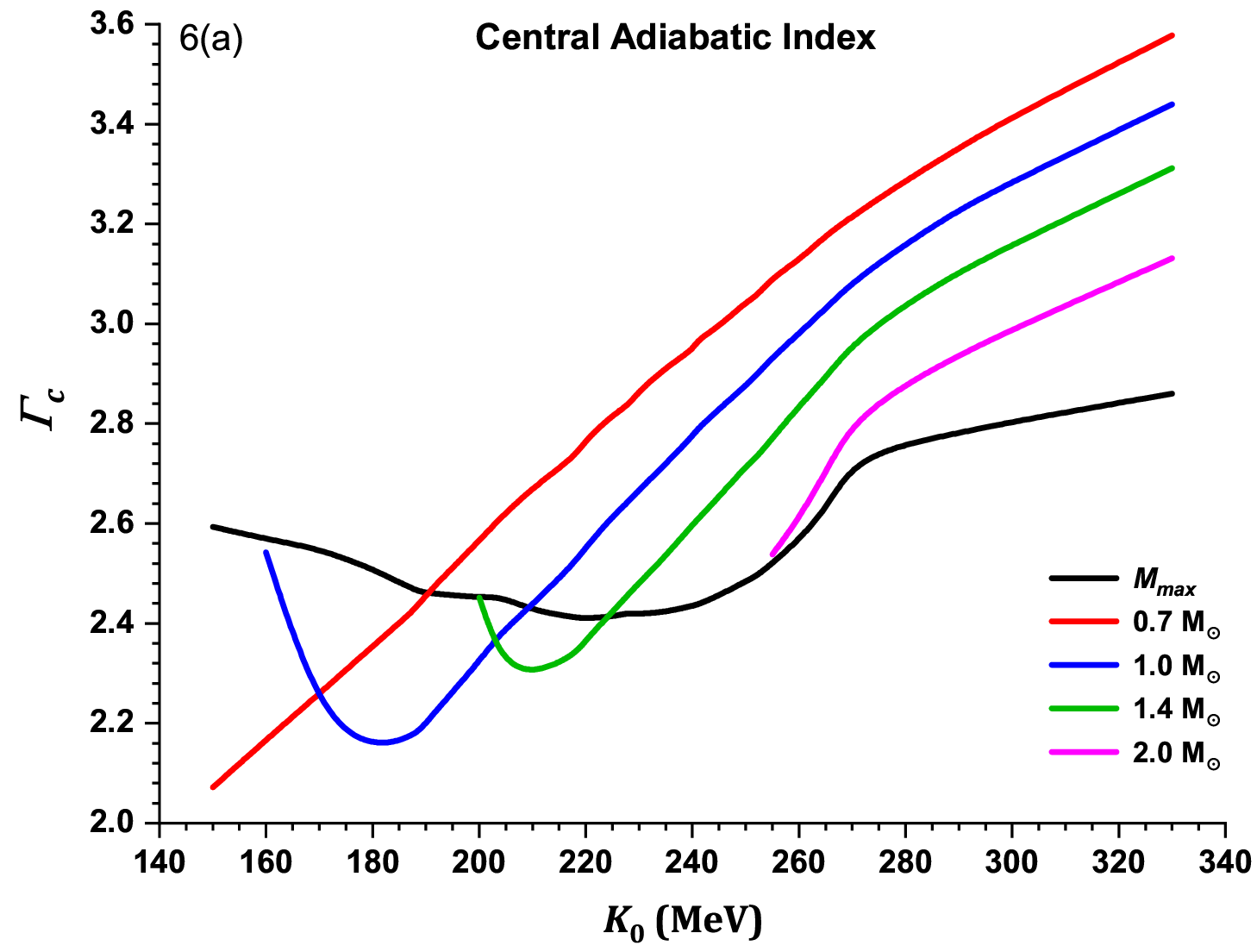}}
	~
	\subfigure[ \label{fig:Fig6b}]{
		\includegraphics[height=12cm,width=1.0\linewidth]{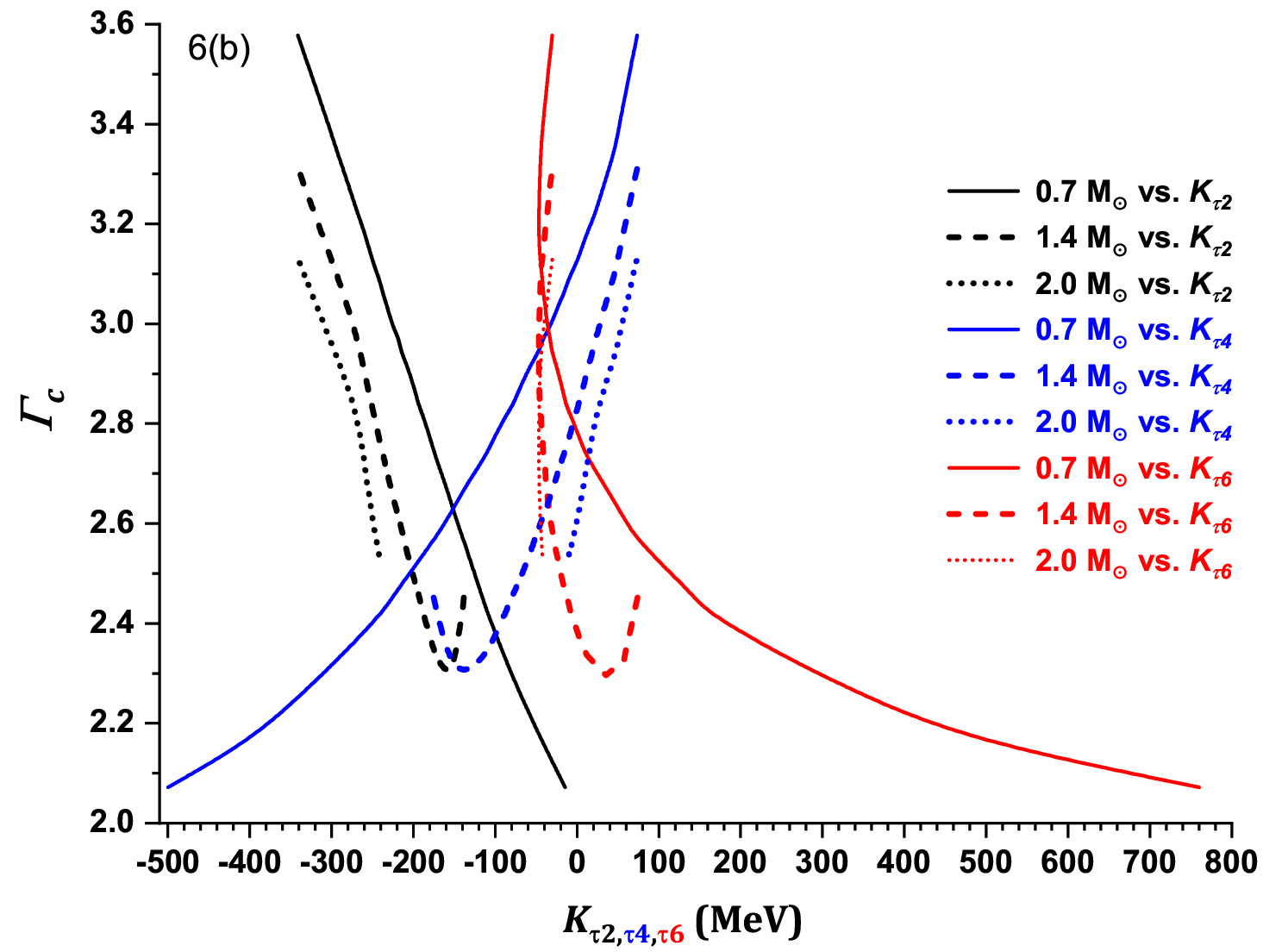}}	
	\caption{\label{fig:Figure6} The adiabatic index at the center of non-rotating NS of the same masses and based on the same EOSs in Fig. \ref{fig:Figure4}, as a function of (a) ${K}_{0}$ and of (b) the isobaric incompressibility coefficients ${K}_{\tau2}$, ${K}_{\tau4}$, and ${K}_{\tau6}$ represented at the same axis.}
\end{figure*}

The final quantity examined for its dependence on ANM stiffness and symmetry-energy coefficients is the thermodynamic adiabatic index, which is one of the parameters critically governing the compact stellar configuration and its thermodynamic stability. Defined as the ratio of specific heats at constant pressure and constant volume, this index reflects the pressure-density variation at constant entropy. The variation of the central adiabatic index (${\varGamma}_{c}$) with the isoscalar incompressibility coefficient is presented in Fig. \ref{fig:Fig6a}, at the center of non-rotating NSs of the masses displayed in Fig. \ref{fig:Fig4a}, far from expected influences of complex structure including nuclear pasta phases that generate non-monotonic $\varGamma$ variations near inner crust. Figure \ref{fig:Fig6a} shows that ${\varGamma}_{c}$ decreases with NS mass, correlated with rising both the central density and the charged particle fractions. For both ${\text{M}}_{\odot}$ and 1.4 ${\text{M}}_{\odot}$ configurations, ${\varGamma}_{c}$ displays a distinct minimum before exhibiting its characteristic positive correlation with ${K}_{0}$, consistent with the general trend observed across the different NS masses. Below this critical point, the soft EOS models show a rapid decline in central pressure with even modest increases in EOS stiffness. The estimated central ${\varGamma}_{c}$ characterizing ${M}_{\text{max}}$(${K}_{0}$) decreases as ${K}_{0}$ increases up to 240 MeV, beyond which it begins to rise. Within the range of ${K}_{0}=240-270$  MeV, the pressure behavior shifts; initially, it increases with growing NM stiffness, but beyond ${K}_{0}$ = 270 MeV, it exhibits a slight downward trend. Increasing the adiabatic index strengthens the NS's resistance to gravitational collapse into a black hole, indicating improved stability \cite{Moustakidis2017,PosadaHladik2021}. \par 
To examine the above-mentioned correlations between symmetry-energy coefficients and NM stiffness, and their influence on the central adiabatic index, Fig. \ref{fig:Fig6b} illustrates how ${\varGamma}_{c}$ varies with the three isovector asymmetry incompressibility coefficients ${K}_{\tau2,\tau4,\tau6}$. Based on the CDM3Y-Paris-IVF1 EOSs over the range ${K}_{0}=150-330$ MeV, ${K}_{\tau2}$ becomes more negative (from $-15$ MeV to $-341$ MeV), ${K}_{\tau4}$ shifts from $-500$ MeV to 73 MeV, and ${K}_{\tau6}$ decreases over more extended range from 760 MeV to $-30$ MeV. As expected, ${\varGamma}_{c}$ increases as the negativity of ${K}_{\tau4}$ decreases, mirroring its trend with ${K}_{0}$, remaining its expected behavior with increasing ${Q}_{4}$ and decreasing the negativity of ${Q}_{0}$, ${H}_{0}$, ${I}_{2,4}$, ${G}_{2}$ and ${K}_{4}$. In contrast, ${\varGamma}_{c}$ shows a decreasing trend with decreasing the negativity of ${K}_{\tau2}$ and increasing ${K}_{\tau6}$, opposite to its trend with ${K}_{0}$ and analogous to its functional behavior with increasing ${I}_{0}$, ${G}_{0}$, ${Q}_{2}$, ${H}_{2}$, and ${K}_{6}$ and with reducing the negativity of ${K}_{2}$. The influence of higher-order symmetry-energy coefficients on ${\varGamma}_{c}$ increases as the NS mass decreases, due to the softer EOSs that successfully describe low-mass NS. \par 
In Table \ref{table1}, we summarize the estimated impacts of the higher order symmetry-energy coefficients on the different investigated quantities in the present work and in our previous studies \cite{SeifHashem2025,SeifHashemJPG2024}. In order to rigorously test and validate these theoretical predictions, and their model dependence, it will be crucial to combine future astrophysical observations spanning multiple messengers with laboratory experiments probing the properties of dense matter under extreme conditions.     
       
\section{\label{sec:level4}SUMMARY AND CONCLUSIONS}
We studied the influence of the EOS of the NS core matter and its higher-order isoscalar and isovector symmetry energy coefficients on its core-crust transition properties, direct Urca threshold conditions, moment of inertia, adiabatic index, and other related quantities. The analysis utilizes EOSs derived within a non-relativistic HF framework using the semi-realistic CDM3Y-Paris-IVF1 nucleon-nucleon interaction, covering a broad stiffness range characterized by ${K}_{0}=150-330$ MeV. We found that the transition pressure and density at the core-crust boundary increase with higher symmetry energy (${E}_{sym0}$) and its corresponding larger density-slope ($L$), but for fixed ${E}_{sym0}$ values, ${P}_{t}$ decreases if $L$ increases. While the stiffness effect strengthens with increasing ${E}_{sym0}$ and $L$, its influence weakens progressively from core-crust transition density and pressure to the corresponding proton fraction. The fractions of $p$, $e$, and $\mu$ decrease monotonically as the stiffness of the core matter increases, correlating with larger indicated ${M}_{\text{max}}$(${K}_{0}$) but less central density. Stiffer NM mildly suppresses $e/p$, and slightly $\mu/p$ and $\mu/e$ ratios. At fixed stiffness, the individual ${x}_{ec}$ and ${x}_{\mu c}$ and ${x}_{\mu c}/{x}_{pc}$ and ${x}_{\mu c}/{x}_{ec}$ slightly increase with the NS mass, while ${x}_{ec}/{x}_{pc}$ slightly decreases. The isovector density dependence predicting stiff high-density symmetry energy yields a lower threshold DU proton fraction, density, and pressure, and indicates allowed DU processes across a wide stiffness range, compared to the softer symmetry energy at supra-saturation densities with less density-slope $L$, which restricts DU to only extremely soft NM. Lower DU thresholds speed up cooling, while higher values limit DU to massive NS. A thicker (thinner) crust corresponds to an earlier (delayed) core-crust transition at higher (lower) densities. Unified linear decreasing dependence of both $I$ and $\Delta I/I$ on the NS compactness is obtained, though expanding distinct ranges of $\mathpzc{C}(M)$ ranges, Ultimately, $I$ rises while $\Delta I/I$ falls with $\mathpzc{C}$ for the estimated ${M}_{\text{max}}$(${K}_{0}$). The compensating EOS effects such as increasing $M$ and $R$ offset by decreasing central density preserve this indicated trend across different EOSs. For stiff EOSs, higher $\mathpzc{C}$ at large $M$ compresses the crust to minimal thickness, yielding remarkably consistent $\Delta I/I$ over wide stiffness range (${K}_{0}>250$ MeV). \par
The present results demonstrate that stiffer NM EOS, marked by larger isoscalar incompressibility (${K}_{0}$) and fourth-order skewness (${Q}_{4}$), along with reduced negativity of isoscalar coefficients ${Q}_{0}$ and ${H}_{0}$, and of the higher-order isovector ${K}_{4}$, ${K}_{\tau4}$, ${I}_{2,4}$, and ${G}_{2}$ coefficients, enhances several NS properties. These include core-crust transition pressure, density, and charged-particle fractions, their direct Urca threshold values, total MI and its crustal fraction, in addition to the central adiabatic index. In contrast, these stiffer EOS conditions reduce the central charged-particle fractions, fractional crust thickness and MI for the maximum NS mass indicated at a given ${K}_{0}$. On the other hand, increasing the isoscalar ${I}_{0}$ and ${G}_{0}$ symmetry coefficients, and higher-order ${Q}_{2}$, ${K}_{6}$, ${K}_{\tau6}$, and ${H}_{2}$ isovector coefficients, accompanied by less negative of ${K}_{2}$ and ${K}_{\tau2}$, which anti-correlate with the NM stiffness decreases the core-crust transition and direct Urca threshold values of pressure, density, and charged-particle fractions, total MI and its crustal fraction, as well as the central adiabatic index. Concurrently, they increase the central non-neutron fractions, fractional crust thickness and MI for the ${M}_{\text{max}}$(${K}_{0}$) configuration. Future astrophysical multimessenger observations and laboratory dense-matter experiments will be essential to validate such predictions.




\begin{table*}[htbp]
	\centering
	\caption{\label{table1}Systematic impact of higher-order symmetry-energy coefficients on observable quantities and properties of NS, as determined in the present work and in our previous studies \cite{SeifHashem2025,SeifHashemJPG2024}, based on the CDM3Y-Paris-IVF1 equations of state characterized by 150 MeV$\le{K}_{0}$(SNM)$\le$ 330 MeV. The investigated symmetry-energy coefficients include the isoscalar (${K}_{0}$) and isovector (${K}_{2,4,6}$ and ${K}_{\tau2,\tau4,\tau6}$) incompressibility, skewness (${Q}_{0,2,4,6}$), kurtosis (${I}_{0,2,4}$), the fifth-order ${H}_{0,2}$, and the sixth-order ${G}_{0,2}$ symmetry-energy coefficients.}
	\begin{ruledtabular}
		\begin{tabular}{lcccc}
			\multirow{7}{*}{\textbf{NS property}} & \multicolumn{4}{c}{\textbf{Higher-order symmetry-energy coefficients}} \\ \cline{2-5}
			&
			\makecell{Increasing with \\the stiffness of \\the EOS} & \makecell{Its negativity \\decreases with\\the stiffness} & \makecell{Decreasing \\with the \\stiffness} & \makecell{Its negativity\\increases with\\the stiffness} \\ \cline{2-5}
			&
			\multirow{2}{*}{$\bm{{K}_{0},\,{Q}_{4}}$} & \multirow{2}{*}{\makecell{$\bm{{Q}_{0},\,{H}_{0},\,{K}_{4},}$\\$\bm{{I}_{2,4},\,{G}_{2},\,{K}_{\tau4}}$}} & \multirow{2}{*}{\makecell{$\bm{{I}_{0},\,{G}_{0},\,{Q}_{2},}$\\$\bm{{K}_{6},\,{H}_{2},\,{K}_{\tau6}}$}} & \multirow{2}{*}{$\bm{{K}_{2}\, \text{\textbf{and}}\,{K}_{\tau2}}$} \\&&&&\\ \hline &&&&\\
			\tabitem Mass (NS)
			 & \multicolumn{2}{c}{\multirow{17}{*}{Increase}}&\multicolumn{2}{c}{\multirow{17}{*}{Decrease}}\\
			\tabitem ${M}_{\text{max}}({K}_{0})$  &&&&\\ 
			\tabitem Radius &&&&\\
			\tabitem Compactness of ${M}_{\text{max}}({K}_{0})$  &&&&\\
			\tabitem Tidal deformability and Love number &&&&\\
			\tabitem Surface redshift of  ${M}_{\text{max}}({K}_{0})$  &&&&\\
			\tabitem Central speed of sound of   ${M}_{\text{max}}({K}_{0})$  &&&&\\
			
			 \makecell[l]{\tabitem Core-crust transition pressure\\~~~~and density, and proton, electron\\~~~~and muon fractions}  &&&&\\
			 \makecell[l]{\tabitem Direct Urca pressure and\\~~~~density, and proton, electron\\~~~~and muon fractions}  &&&&\\
			\tabitem Moment of inertia &&&&\\
			\makecell[l]{\tabitem Fractional crust thickness\\~~~~and moment of inertia}  &&&&\\
			\tabitem Central adiabatic index &&&&\\
			\hline
			&&&&\\
			\tabitem Compactness & \multicolumn{2}{c}{\multirow{11}{*}{Decrease}}&\multicolumn{2}{c}{\multirow{11}{*}{Increase}}\\
			\makecell[l]{\tabitem Tidal deformability and\\~~~~Love number of ${M}_{\text{max}}({K}_{0})$}  &&&&\\
			\tabitem Surface redshift &&&&\\
			\tabitem Central speed of sound &&&&\\
			\makecell[l]{\tabitem Central proton, electron\\~~~~and muon fractions}  &&&&\\
			\makecell[l]{\tabitem Fractional crust thickness and\\~~~~moment of inertia of ${M}_{\text{max}}({K}_{0})$}  &&&&\\
			\tabitem Central energy-density and pressure &&&&\\
			\tabitem Central density &&&&\\

	     \end{tabular}
    \end{ruledtabular}
\end{table*}	

\end{document}